\documentclass[]{aastex631}

\usepackage{subfigure}
\usepackage{longtable}

\graphicspath{{./}{figures/}}

\begin{document}

\title{A Study of Radio Knots within Supernova Remnant Cassiopeia A}

\author[0000-0002-5223-3653]{Xianhuan Lei}
\affiliation{National Astronomical Observatories, Chinese Academy of Sciences, 20A Datun Road, Chaoyang District, Beijing 100012, China; zhuhui@bao.ac.cn, tww@bao.ac.cn, hyzhang@nao.cas.cn}
\affiliation{School of Astronomy and Space Science, University of Chinese Academy of Sciences, Beijing 100049, China}

\author{Hui Zhu}
\affiliation{National Astronomical Observatories, Chinese Academy of Sciences, 20A Datun Road, Chaoyang District, Beijing 100012, China; zhuhui@bao.ac.cn, tww@bao.ac.cn, hyzhang@nao.cas.cn}

\author{Haiyan Zhang}
\affiliation{National Astronomical Observatories, Chinese Academy of Sciences, 20A Datun Road, Chaoyang District, Beijing 100012, China; zhuhui@bao.ac.cn, tww@bao.ac.cn, hyzhang@nao.cas.cn}
\affiliation{CAS Key Laboratory of FAST, National Astronomical Observatories, Chinese Academy of Sciences, Beijing 100101, China}

\author{Wenwu Tian}
\affiliation{National Astronomical Observatories, Chinese Academy of Sciences, 20A Datun Road, Chaoyang District, Beijing 100012, China; zhuhui@bao.ac.cn, tww@bao.ac.cn, hyzhang@nao.cas.cn}
\affiliation{School of Astronomy and Space Science, University of Chinese Academy of Sciences, Beijing 100049, China}
\affiliation{The Key Laboratory of Cosmic Rays (Tibet University), Ministry of Education, Lhasa 850000, China}

\author{Dan Wu}
\affiliation{National Astronomical Observatories, Chinese Academy of Sciences, 20A Datun Road, Chaoyang District, Beijing 100012, China; zhuhui@bao.ac.cn, tww@bao.ac.cn, hyzhang@nao.cas.cn}


\begin{abstract}

The study on the dynamic evolution of young supernova remnants (SNRs) is an important way to understand the density structure of the progenitor's circumstellar medium. We have reported the acceleration or deceleration, proper motion and brightness changes of 260 compact radio features in the second youngest known SNR Cas A at 5\,GHz based on the VLA data of five epochs from 1987 to 2004. The radio expansion center locates at $\alpha(1950)=23^{\rm h}21^{\rm m}9^{\rm s}_{\cdot}7 \pm 0^{\rm s}_{\cdot}29, \delta(1950)=+58^{\circ}32^{\prime}25^{\prime\prime}_{\cdot}2 \pm 2^{\prime\prime}_{\cdot}2$. Three-quarters of the compact knots are decelerating, this suggests that there are significant density fluctuations in the stellar winds of the remnant's progenitor. We have verified that the acceleration or deceleration of compact knots are not related with the distribution of brightness. The brightening, fading, disappearing or new appearing of compact radio features in Cas A suggests that the magnetic field in the remnant is changing rapidly.

\end{abstract}


\section{Introduction}
Supernova remnants (SNRs) are the result of supernova (SN) ejecta interaction with the interstellar medium (ISM). The density structure of the ISM can be revealed by the early dynamic evolution of young SNRs (\citealp{2012ApJ...744...71I}; \citealp{2021Ap&SS.366...58S}). Cassiopeia A (Cas A; 111.7-2.1), with the expansion date estimated around A.D. 1680 (\citealp{1980JHA....11....1A}; \citealp{2006ApJ...645..283F}), is the second-youngest known Galactic SNR, at a distance of $\sim$3.4\,kpc \citep{1995ApJ...440..706R}. The observed light echoes in Cas A suggests that this remnant came from a Type IIb SN explosion \citep{2008Sci...320.1195K}. With the observations from the longest radio wave to very high energy $\gamma$-ray ( \citealp{1975Natur.257..463B}; \citealp{2004ApJ...614..727M}; \citealp{2001ApJ...552L..39G}; \citealp{2017MNRAS.472.2956A}), Cas A is one of the most-studied objects in the sky.

Multi-wavelength kinematics analysis of fragmented filaments or knots in shock provides important information about the radiative and dynamical evolution of Cas A. Generally, bright optical knots and X-ray knots of Cas A are not consistent due to the differences of temperature and density \citep{2014ApJ...789..138P}, the size of X-ray knots ($\sim1.0^{\prime\prime}-5.0^{\prime\prime}$: \citealp{2007AJ....133..147P}) appears to be 10 times of optical knots ($\sim0.2^{\prime\prime}-0.3^{\prime\prime}$: \citealp{2001AJ....122.2644F}). In radio bands, the shell of Cas A has two obvious features, a bright ring with a radius of $\sim$1.7 pc, and a faint plateau extending to a radius of $\sim$2.5 pc \citep{2014ApJ...785..130Z}. The radio knots mainly distribute in the two regions, and cover a variety of spatial scales \citep{1986MNRAS.219...13T}.

Difference in spectroscopy and velocities has divided the optical emission in Cas A into three distinct types: 1) fast-moving knots (FMKs) with velocities from 4000 $\rm km\,\rm s^{-1}$ to 6000 $\rm km\,\rm s^{-1}$ (\citealp{1988ApJ...329L..89F}; \citealp{2001AJ....122..297T}), which have strong O, Si, S, Ar, Ca emission line but no H or He emission (\citealp{{1978ApJ...219..931C},{1979ApJ...233..154C}}); 2) quasi-stationary flocculi (QSF) with velocities less than 500 $\rm km\,\rm s^{-1}$ \citep{1988ApJ...329L..89F} and exhibit H, He, N emission (\citealp{{1978ApJ...219..931C}, {1979ApJ...233..154C}}); 3) fast-moving flocculi (FMF), which is the outlying faint emission knots of the remnant with velocities between 6500 $\rm km\,\rm s^{-1}$ and 8600 km $\rm s^{-1}$ (\citealp{{1987ApJ...313..378F}, {1988ApJ...329L..89F}}), and have H, S and N emission \citep{2001ApJS..133..161F}.

Similar to optical emission, the X-ray spectrum with O, Si, S, Ar, Ni, Ca and Fe emission lines, and the spatial distribution of element abundances in X-ray ejecta reveals the nucleosynthesis processes of SN explosion (\citealp{2000ApJ...528L.109H}; \citealp{2002A&A...381.1039W}; \citealp{2004ApJ...613..343D}; \citealp{2012ApJ...746..130H}). Non-thermal radiation extends to hard X-rays in both the forward and reverse shock (\citealp{1997ApJ...487L..97A}; \citealp{2015ApJ...802...15G}). The non-thermal faint X-ray filaments or knots located at the edge of Cas A and thus identifies the forward shock whose expansion velocity has been measured in the range of $\sim$4000 km $\rm s^{-1}$ to 6000 km $\rm s^{-1}$ (\citealp{2001ApJ...552L..39G}; \citealp{2003ApJ...589..818D}; \citealp{2009ApJ...697..535P};  Vink et al. \citealp{2022ApJ...929...57V}). \citet{2008ApJ...686.1094H} had shown that most of non-thermal X-ray emission comes from fragmented filaments  in reverse-shock, with the reverse-shock showed obvious asymmetry \citep{2008ApJ...677L.105U}. The reverse-shock ejecta has an expansion velocity of $\sim$5000 km $\rm s^{-1}$, while the shock velocity at west reached to 6000 km $\rm s^{-1}$ (\citealp{2007AJ....133..147P}; \citealp{2009PASJ...61.1217M}; \citealp{2018ApJ...853...46S}; \citealp{2022A&A...666A...2O}).

High-resolution observations of Cas A made by aperture synthetic systems provide important approach for detection the evolution of radio emission and the detailed sampling of velocity field. \citet{1977MNRAS.179..573B} compared two images taken by Cambridge Telescope at 5 GHz in 1969 and 1974 (\citealp{1970MNRAS.151..109R}; \citealp{1975Natur.257..463B}), the flux changes and proper motion of 34 compact radio peaks were measured. Combined with the observations of NRAO interferometer at 8.19 GHz in 1976 and  2.7 GHz in 1967, the evolution of 83 radio emission features had been studied, of which $\sim$30\% were brightening (\citealp{1979A&A....75...44D}; \citealp{1969AJ.....74.1206H}). \citet{1986MNRAS.219...13T} made an observation by Cambridge 5-km Telescope at 5 GHz in 1978 and compared the observation made by \citet{1975Natur.257..463B} in 1974, measured proper motion and brightness changes of 342 compact radio features. There was no relationship between proper motion and brightness change. \citeauthor{1995ApJ...441..307A} (\citeyear{1995ApJ...441..307A}: AR95) calculated the proper motion and brightness changes of 304 radio knots using the least-square method, which combined observations of Cambridge 5-km Telescope and Very Large Array (VLA\footnote{\url{https://science.nrao.edu/facilities/vla}}) at 5 GHz for 6 epochs (Cambridge 5-km Telescope: 1978, 1982; VLA: 1983, 1985, 1987, 1990). The radio emission plasma in Cas A was decelerating significantly, and the bulk expansion age was 2.5--4 times the actual age of the remnant. Brighter knots appear to be steeper spectra, and do not find the correlation between spectral index and brightness changes \citep{1996ApJ...456..234A}.

In this work, we report the progenitor's circumstellar medium density distribution of Cas A by studying the motion and brightness changes of the radio emissions in the remnant. We use the VLA archival visibility observation data for Cas A at 5 GHz including five epochs of 1987, 1994, 1997, 2000 and 2004. These data were observed in the A-, B-, C- and D-configuration of VLA, which has a high-sensitivity and high-resolution of $1.5^{\prime\prime}$. For this paper, Section \ref{data} describes the data used and the data calibration process, Section \ref{method} is the calculation for the proper motion and brightness changes of compact radio peaks, the results and discussion are shown in the Section \ref{result}.

\begin{deluxetable*}{cccccc}
\tablenum{1}
\tablecaption{Summary of VLA observation for Cas A from 1987 to 2005, $\lambda=6$ cm.\label{table_1}}
\tablewidth{0pt}
\tablehead{
\colhead{Date} & \colhead{VLA} & \colhead{Frequencies} & \colhead{Bandwidth} & \colhead{On-Source Time} & \colhead{Observer}\\
\colhead{year} & \colhead{configuration} & \colhead{GHz} & \colhead{MHz} & \colhead{minutes} & \colhead{}
}
\startdata
\multicolumn{6}{c}{AB0434(1987--1988)}\\
\hline
1987.58 & A & 4.64, 4.97 & 3.125 & 420 & L.Rudnick \\
1987.62 & A & 4.41, 5.08 & 3.125 & 325 & L.Rudnick \\
1987.77 & A & 4.97, 4.64, 4.41, 5.08 & 3.125 & 330 & L.Rudnick \\
1987.88 & B & 4.97, 4.64 & 6.25 & 357 & L.Rudnick \\
1988.19 & C, D & 4.87, 4.82 & 12.5 & 53 & L.Rudnick \\
1987.35 & D & 4.89, 4.84 & 50 & 30 & L.Rudnick \\
\hline
\multicolumn{6}{c}{AR0310(1994--1995)}\\
\hline
1994.23 & A & 4.41, 4.64, 4.97, 5.08, 4.97 & 3.125 & 545 & L.Rudnick \\
1994.58 & B & 4.41, 4.99, 4.64, 5.09 & 12.5 & 283 & L.Rudnick \\
1994.96 & C & 4.41, 4.99, 4.64, 5.09 & 12.5 & 182 & L.Rudnick \\
1995.23 & D & 4.41, 4.99, 4.64, 5.08 & 12.5 & 108 & L.Rudnick
 \\
\hline
\multicolumn{6}{c}{AR0378(1997--1998)}\\
\hline
1998.22 & A & 4.41, 4.99, 4.64, 5.08 & 6.25 & 453 & L.Rudnick \\
1997.38 & B & 4.41, 4.99, 4.64, 5.08 & 12.5 & 304 & L.Rudnick \& B.Koralesky \\
1997.70 & C & 4.41, 4.99, 4.64, 5.08 & 12.5 & 142 & L.Rudnick \& B.Koralesky \\
1997.36 & D & 4.41, 4.99, 4.64, 5.08 & 12.5 & 123 & L.Rudnick \\
\hline
\multicolumn{6}{c}{AR0435(2000--2001)}\\
\hline
2000.94 & A & 4.72, 5.0, 4.86, 4.61 & 6.25 & 537 & L.Rudnick \\
2001.33 & B & 4.72, 4.99, 4.86, 4.6 & 12.5 & 468 & L.Rudnick \\
2000.31 & C & 4.72, 5.0, 4.6, 4.86 & 25 & 241 & L.Rudnick \\
2000.68 & D & 4.72, 5.0, 4.86, 4.6 & 12.5 & 183 & L.Rudnick \\
\hline
\multicolumn{6}{c}{AD0500(2004--2005)}\\
\hline
2004.73 & A & 4.72, 4.99, 4.86, 4.61 & 6.25 & 481 & T.Delaney \\
2005.35 & B & 4.86, 4.61, 4.72, 5.0 & 12.5 & 334 & T.Delaney \\
2004.24 & C & 4.72, 5.0, 4.86, 4.6 & 25 & 225 & T.Delaney \\
2004.62 & D & 4.71, 4.99, 4.86, 4.61 & 50 & 163 & T.Delaney \\
\enddata
\end{deluxetable*}

\section{Data Sample and Calibration \label{data}}
The data used in this study is the VLA archival\footnote{\url{https://archive.nrao.edu/archive/archiveproject.jsp}} C-band (5 GHz; $\lambda$=6 cm) standard-configuration data of Cas A. Each data was obtained at several frequencies around the nominal observing frequency of 5 GHz, which improves the aperture coverage. In the VLA standard-configurations, the A-configuration is the most diffuse, with a maximum baseline of 36.4 km and a resolution of 0.33 arcsec in C-Band; D-configuration is the most compact, with the shortest baseline of 0.035 km. The details of the data we used are summarized in Table \ref{table_1}, which covers 5 epochs of 1987.73 (Project: AB0434), 1994.75 (Project: AR0310), 1997.67 (Project: AR0378), 2000.82 (Project: AR0435) and 2004.74 (Project: AD0500), with total time baseline spanning about 17 yrs. In order to sample the complex spatial structure of Cas A, the aperture coverage has been maximized.

These data calibration procedure has performed by the Common Astronomy Software Applications package (CASA\footnote{\url{https://casa.nrao.edu/}}: \citealp{2007ASPC..376..127M}). The calibrations of the primary flux density in all observing sessions are based on the 0134+329 (3C48), and the 2352+4995 is the phase calibrators. The antenna position was first calibrated, then we determined the flux density scale, complex gain, and baseline-based solutions from the calibrator 3C48, and applied them to the target Cas A (3C461). Data for each array were calibrated individually, and then concatenated and imaged. To avoid negative "bowl" around the source, Multi-Scale Clean deconvolution method was used for imaging \citep{2008ISTSP...2..793C}. We have used briggs weighting with different parameters of robust=-2, -1, 0, 1, 2 to restore images. When robust=1 and 2, images are in high random noise of VLA, and rms noise reaches 30 mJy $\rm beam^{-1}$. In the absolute flux scale uncertainty of 3\% \citep{2013ApJS..204...19P}, there are same integrated flux density and stable noise levels for robust=-2, -1 and 0. Considering surface brightness sensitivity while requiring high-resolution, we used robust=0 for the images restore, obtained images with a rms noise of 3.0 mJy $\rm beam^{-1}$.

\section{Method \label{method}}
AR95 has performed a detailed study for the temporal evolution of compact radio features in Cas A. The net radial deceleration of compact radio features in Cas A has not been observed. In this section, we have made a new measurement of the proper motion and brightness changes for the radio features based on the higher resolution ($1.5^{\prime\prime}$) and sensitivity observations of VLA.

\subsection{Image Normalization}
The five images have been obtained from difference observation projects over five epochs. These observations exist differences in absolute flux calibration and aperture coverage, which leads to the differences of these five images' reconstruction. In order to minimize the effect of different aperture coverage for imaging, image normalization is necessary. In the five images, AR310 (1994.75 epoch) has the highest resolution ($0.56^{\prime\prime}\times0.46^{\prime\prime}$) and AD500 (2004.74 epoch) has the lowest resolution ($1.42^{\prime\prime}\times1.2^{\prime\prime}$). To compare the flux density of different observations safely, we adopt the gaussian convolution algorithm to smooth all images into the same beamsize of $1.5^{\prime\prime}$.

\begin{figure*}
\centering
\subfigure{
\includegraphics[width=10.8cm]{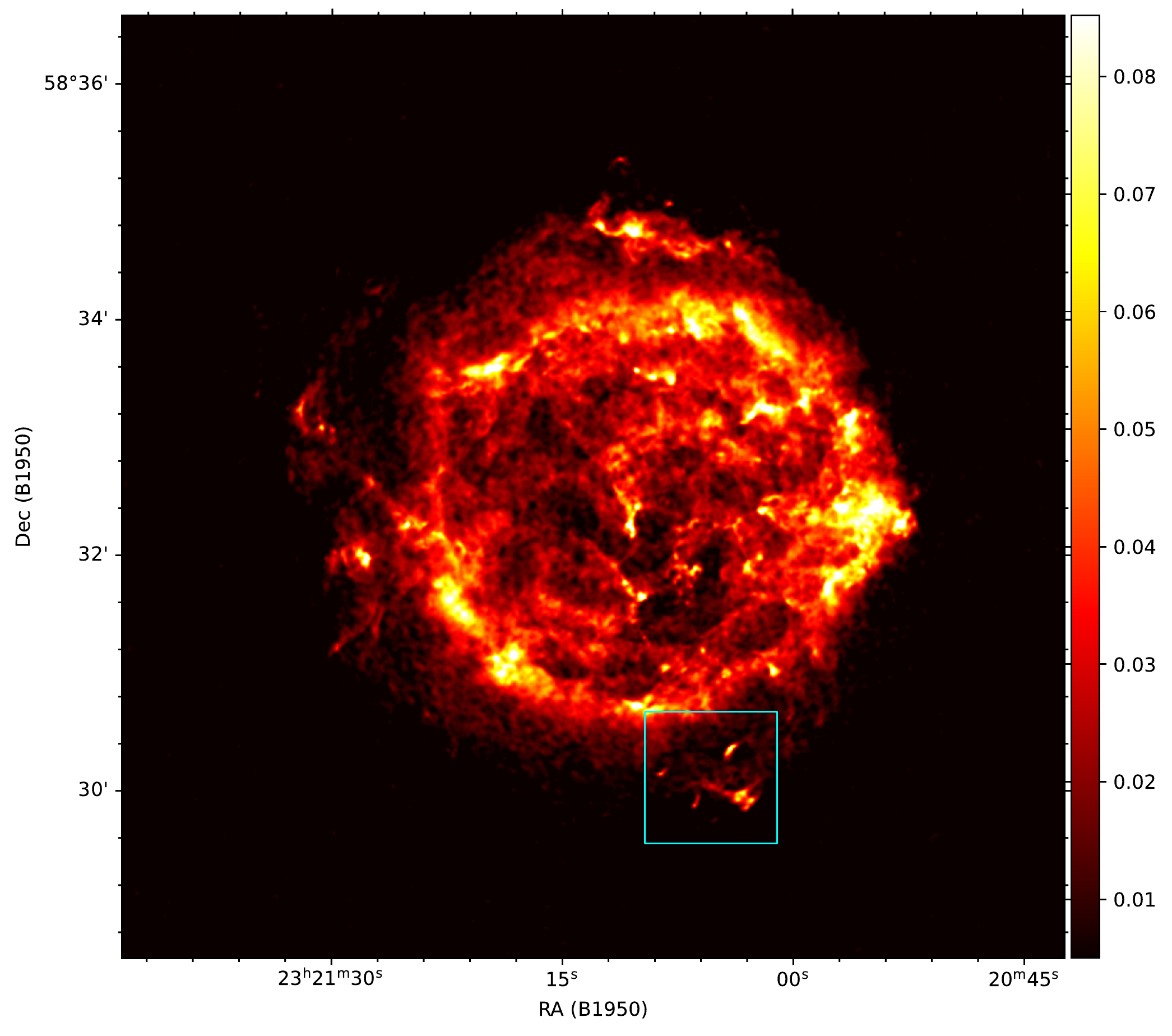}
}\hspace{-4mm}
\quad
\subfigure{
\includegraphics[width=6.7cm]{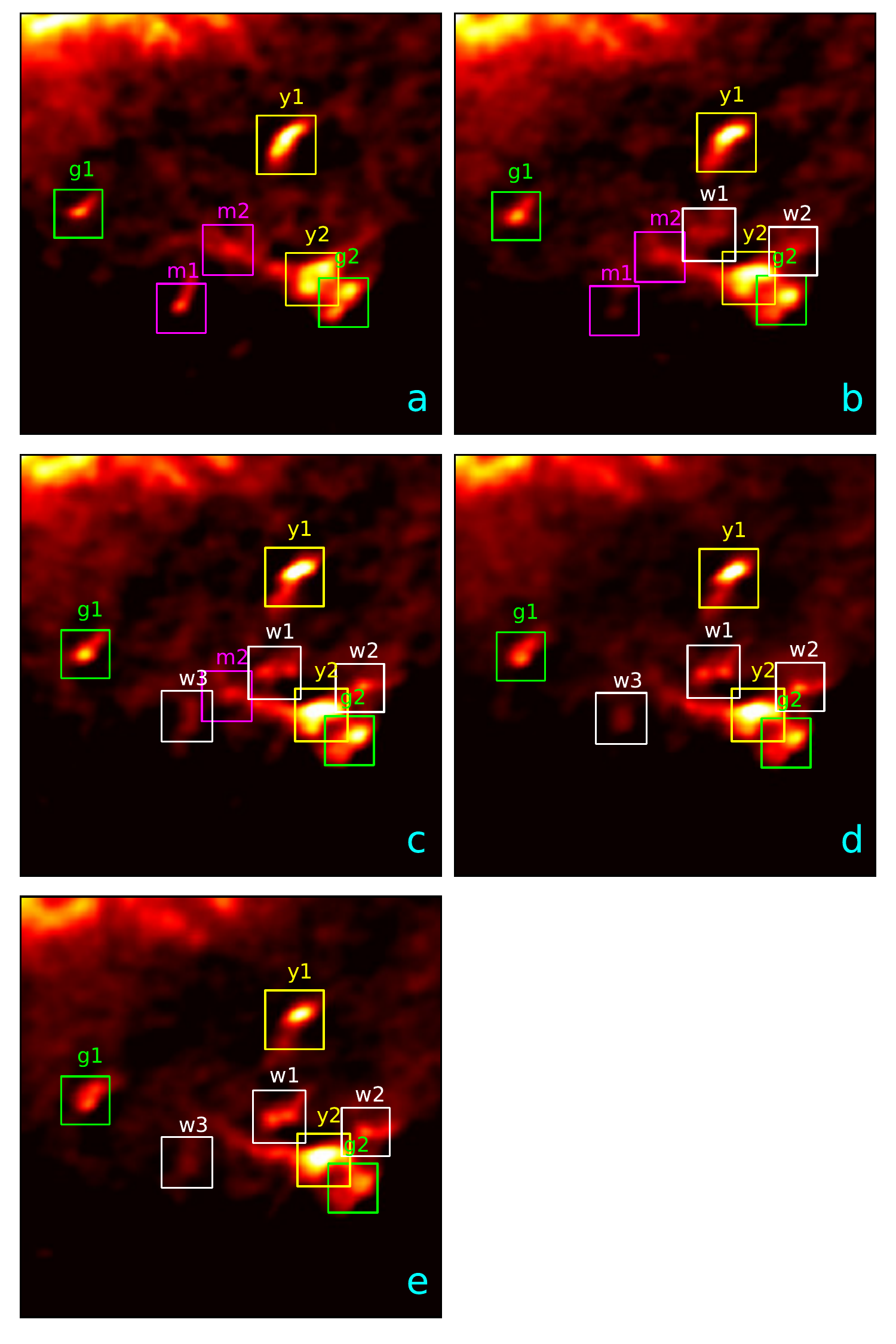}
}
\caption{The continuum images of Cas A at $\lambda=6$
cm, Color-scale is in units of Jy $\rm beam^{-1}$. Left panel: the image of 1987, which combines the A, B, C and D-configuration of VLA. Right panel: the local evolution of Cas A from 1987 to 2004 (the cyan box region on the left panel). Local images of a, b, c, d and e are the 1987 (Project: AB0434), 1994 (Project: AR0310), 1997 (Project: AR0378), 2000 (Project: AR0435) and 2004 (Project: AD0500), respectively. Using the earliest observed image of 1987 as the reference image, the knots in yellow boxes indicates that the knots currently brightening, examples y1 (knot 7) and y2 (knot 5) are brightening from 1987 to 2004. Knots in the green boxes indicates the knots is fading, g1 (knot 3) and g2 (knot 6) are fading from 1987 to 2004. A knot in magenta box will disappear, such as m1 evolves to 1997 and m2 to 2000. In the white boxes shows newly knots, like w1 and w2 appearing in 1994 and w3 appearing in 1997.\label{CasA}}
\end{figure*}

\subsection{Selection of Compact Radio Emission Features}
\begin{figure*}
\centering
\includegraphics[width=17cm]{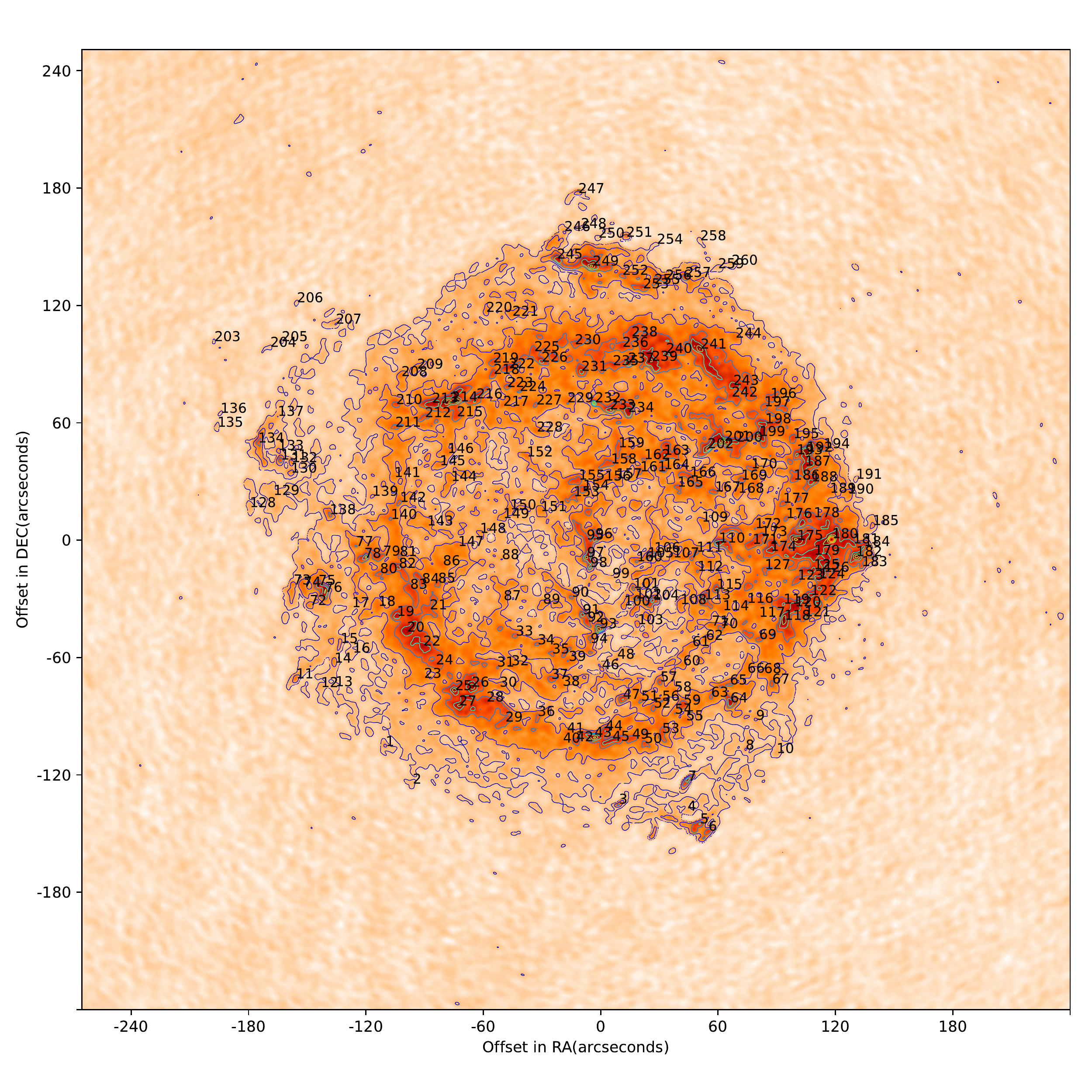}
\caption{Position of the compact radio knots in Cas A constructed from the 1987 5 GHz observations. Contour levels in Jy $\rm beam^{-1}$ are 0, 0.002, 0.004, 0.006, 0.008, 0.010, 0.025, 0.05, 0.075, 0.1, 0.12, 0.14, 0.15, 0.16, 0.18, 0.19 and 0.20.\label{knots_distribution}}
\end{figure*}

Hundreds of compact radio emission features in the remnant of Cas A are evolving rapidly, and parts of the knots are disappearing. To select the radio knots, we denote the location of all local maximum values based on the 1987 image, and construct the contour map. Therefore, we obtain all possible knots in the 1987 image. These knots evolve rapidly, as shown in  Figure \ref{CasA}. In this study, we keep the knots having the following features: it can be isolated from the surroundings; clearly visible in the five continuous images of C-Band; its shape has not changed significantly over the span of 17 yrs. As a result, 260 radio knots have been kept, parts of them are consistent with the 304 knots compiled in AR95. The position of these knots are shows in Figure \ref{knots_distribution}, and the detailed statistics are given in Table \ref{table_2}.

\subsection{Measurement of Position and Brightness Relative Change \label{measurement_position}}
For the position and brightness change of compact features, the least-squares method is widely used (\citealp{1986MNRAS.219...13T}; \citealp{1995ApJ...441..307A}). This numerical method allows us to test acceleration or deceleration of individual knots, which can minimize the effect of different apertures coverage at multi-epochs on the variations of large-scale background reconstruction. We have adopted the same algorithm to measure the positions and brightness changes of each radio knot from epoch 1987 to 2004.

First, we use the epoch 1987 image(the earliest image we used, and have a high resolution) as the reference image $A_{ij}$, the other as comparison images $B_{ij}$ (images from epochs 1994, 1997, 2000 and 2004). A small region is set around each knot of the reference image $A_{ij}$. Next, we select a region for each corresponding knots in comparison images $B_{ij}$, which would contain knot for all epochs and is separated from the surrounding emission. Finally, we minimize the quantity in the region by the following method:
\begin{equation}
T(\xi)=\sum_{\rm{i=1}}^{\rm{I}}\sum_{\rm{j=1}}^{\rm{J}}Q_{ij}^2(\xi),
\end{equation}
\\
the summation T($\xi$) is performed over all pixels in the region(I×J). Where, $\rm Q_{ij}(\xi)={\rm A_{ij}}-{\rm SB_{ij}}(\Delta \alpha,\Delta \sigma)-{\rm H}$, with respect to 4-vector $\xi=(\Delta\alpha, \Delta\sigma$, S, H). S is the brightness scaling factor between different epochs, $\Delta\alpha$ and $\Delta\sigma$ are the variations of the relative position between image $\rm A_{ij}$ and $\rm B_{ij}$, H is the level of change in the local backgrounds. This summation procedure was performed four times at each radio knots to obtain the relative changes of position and brightness in four comparison images and reference image.

\begin{longrotatetable}
\begin{deluxetable*}{ccccccccccccccccccc}
\tablenum{2}
\tablecaption{Summary of VLA observations for Cas A from 1987 to 2005, $\lambda=6$ cm.\label{table_2}}
\tablewidth{700pt}
\tabletypesize{\scriptsize}
\tablehead{
\colhead{knots} & \colhead{$\rm knot_{\rm A95}$} & \colhead{x} & \colhead{y} & \colhead{$\rm Box_{x1}$} & \colhead{$\rm Box_{x2}$} & \colhead{$\rm Box_{y1}$} & \colhead{$\rm Box_{y2}$} & \colhead{$v_x$} & \colhead{$\delta v_x$} & \colhead{$v_y$} & \colhead{$\delta v_y$} & \colhead{$a_x$} & \colhead{$\delta a_x$}  & \colhead{$a_y$} & \colhead{$\delta a_y$} & \colhead{Peak Brightness} & \colhead{$\frac{\Delta B}{B}$} & \colhead{$\delta \frac{\Delta B}{B}$}
}
\decimalcolnumbers
\startdata
1  &  22  &  -97.1  &  -109.7  &  -103.1  &  -115.9  &  -92.0  &  -105.4  &  -0.069  &  0.04  &  -0.177  &  0.01  &  0.0014  &  0.0012  &  0.0026  &  0.0031  &  28.53  &  0.087  &  0.034  \\
2  &  2  &  -83.2  &  -129.0  &  -88.0  &  -134.9  &  -78.8  &  -124.7  &  -0.092  &  0.006  &  -0.221  &  0.04  &  -0.0001  &  0.0013  &  0.0021  &  0.0021  &  17.13  &  0.177  &  0.031  \\
3  &  6  &  22.3  &  -139.3  &  16.7  &  -146.4  &  28.5  &  -135.1  &  0.019  &  0.016  &  -0.101  &  0.009  &  0.0014  &  0.0014  &  0.0002  &  0.0018  &  54.54  &  0.068  &  0.017  \\
4  & \nodata &  57.3  &  -142.7  &  54.5  &  -147.7  &  60.3  &  -139.9  &  -0.042  &  0.019  &  -0.005  &  0.022  &  0.0013  &  0.0014  &  -0.0038  &  0.0012  &  21.2  &  0.13  &  0.018  \\
5  &  11  &  63.7  &  -149.4  &  59.1  &  -155.5  &  68.0  &  -146.5  &  -0.264  &  0.072  &  -0.103  &  0.006  &  0.0128  &  0.0025  &  0.0006  &  0.0014  &  77.13  &  0.069  &  0.01  \\
6  &  13  &  68.0  &  -153.2  &  64.1  &  -158.6  &  74.4  &  -150.4  &  0.126  &  0.016  &  -0.173  &  0.016  &  -0.0008  &  0.0026  &  0.0022  &  0.003  &  69.28  &  0.036  &  0.01  \\
7  &  9  &  57.5  &  -127.5  &  50.8  &  -133.8  &  65.5  &  -121.3  &  0.137  &  0.024  &  0.042  &  0.014  &  -0.0028  &  0.0016  &  -0.0026  &  0.0009  &  87.47  &  0.039  &  0.011  \\
8  &  73  &  87.0  &  -111.5  &  84.1  &  -116.0  &  94.6  &  -105.9  &  0.312  &  0.051  &  0.001  &  0.021  &  -0.0132  &  0.003  &  -0.0037  &  0.0016  &  27.13  &  0.078  &  0.034  \\
\enddata
\tablecomments{Column 1: Number of each knot in this article; Column 2: Number of each knot same as in AR95; Column:3 and 4: Position of compact knots in the epoch 1987 continuum map, measured in arcsec and reference to the optical expansion center $\alpha(1950)=23^{\rm h}21^{\rm m}11^{\rm s}_{\cdot}4$, $\delta(1950)=+58^{\circ}32^{\prime}28^{\prime\prime}_{\cdot}5$; Column 5 to 8: Box size defined for each knot, $x_1,y_1,x_2,y_2$ (arcsec), reference to the radio expansion center; Column 9: $v_x$, the proper motion of each knot in right ascension (arcsec $\rm yr^{-1}$); Column 10: $\delta v_x$, uncertainty in right ascension proper motion; Column 11: $v_y$, the proper motion of each knot in declination (arcsec $\rm yr^{-1}$); Column 12: $\delta v_y$, uncertainty in declination proper motion; Column 13: $a_x$, the acceleration of each knot in right ascension (arcsec $\rm yr^{-2}$); Column 14: $\delta a_x$, uncertainty in right ascension acceleration; Column 15: $a_y$, the acceleration of each knot in declination (arcsec $\rm yr^{-2}$); Column 16: $\delta a_y$, uncertainty in declination acceleration; Column 17: Peak brightness of each knot in 1987(mJy $\rm beam^{-1}$); Column 18: $\frac{\Delta B}{B}$, annual percentage of the brightness change(percent $\rm yr^{-1}$); Column 19: $\delta \frac{\Delta B}{B}$, uncertainty in fractional brightness change.}
\end{deluxetable*}
\end{longrotatetable}

\section{Results and Discussion \label{result}}
The position ($x, y$) and peak brightness of all radio knot samples we studied are given in Table \ref{table_2}. The position respects to the optical expansion center of $\alpha(1950)=23^{\rm h}21^{\rm m}11^{\rm s}_{\cdot}4$, $\delta(1950)=+58^{\circ}32^{\prime}28^{\prime\prime}_{\cdot}5$ \citep{1998ApJ...505L..27K}. The quoted peak brightness is the maximum of the radio emission features. To determine the proper motion, acceleration or deceleration and brightness change ratio, nonlinear fitting method has been used. Proper motion ($v_x, v_y$), acceleration or deceleration ($a_x, a_y$), and brightness change ratio ($\frac{\Delta B}{B}$) fitting results are summarized in Table \ref{table_2}.

\subsection{Proper Motions of Compact Radio Knots \label{motion_result}}
\begin{figure}
\centering
\subfigure{
\includegraphics[width=8.5cm]{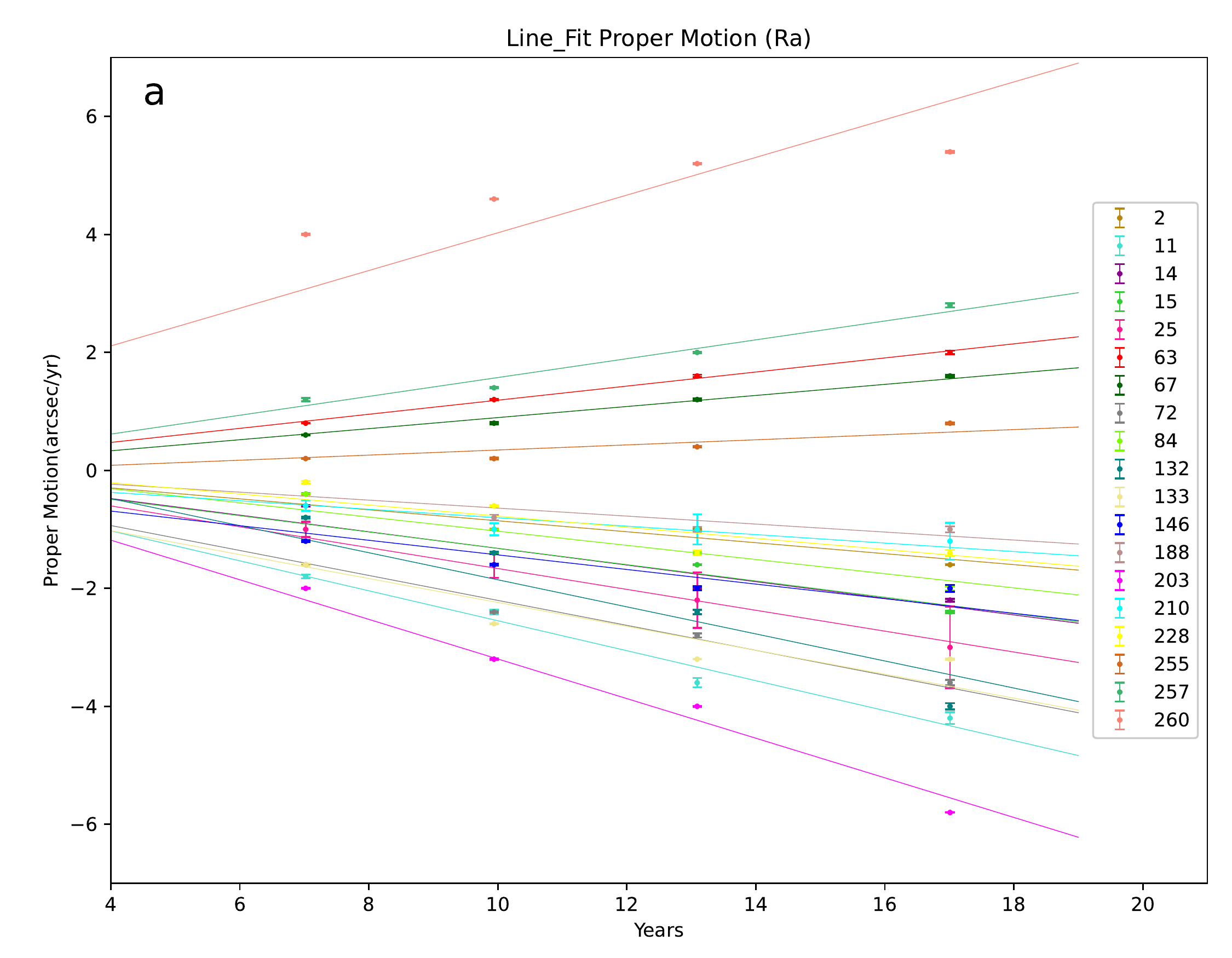}
}
\quad
\subfigure{
\includegraphics[width=8.5cm]{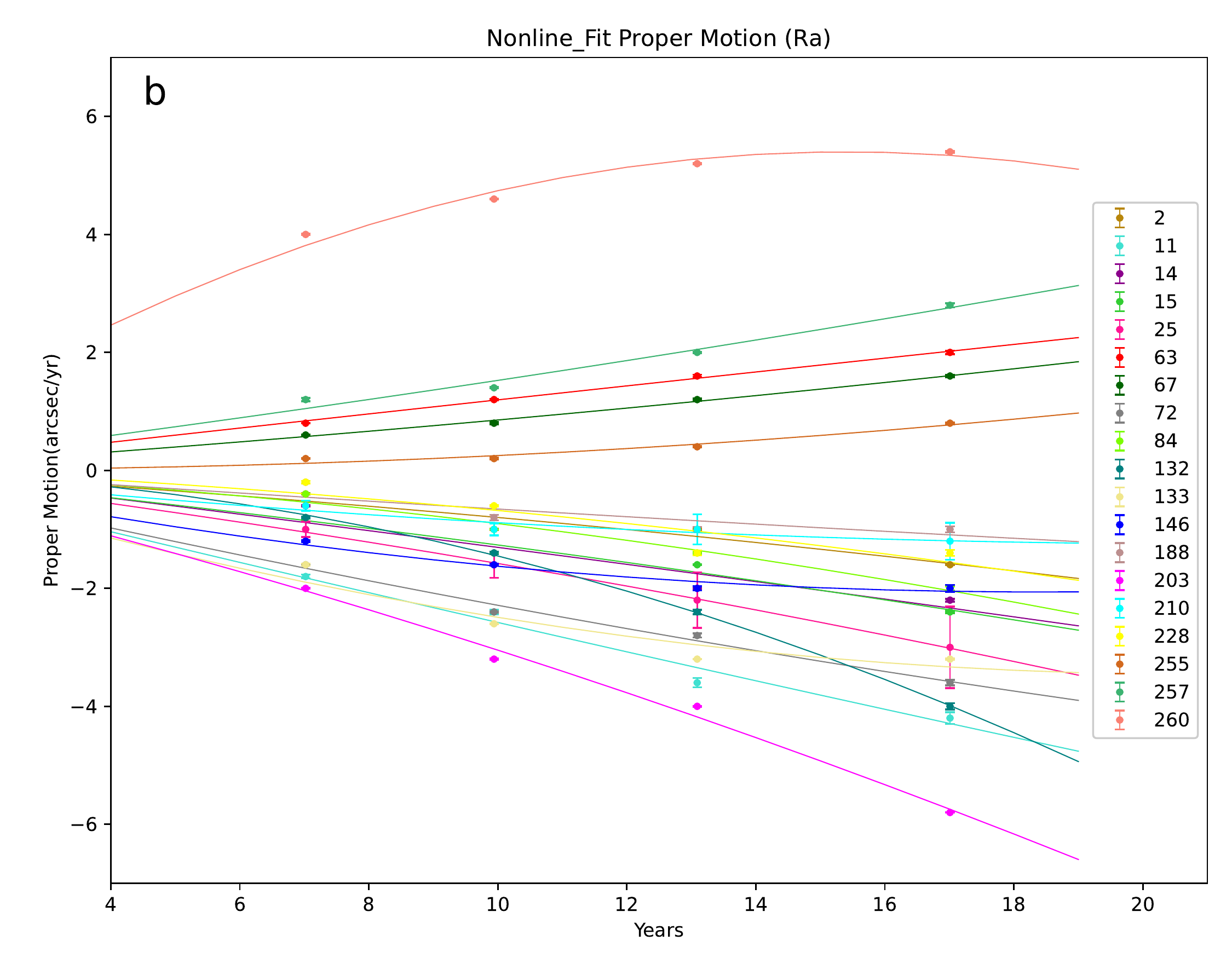}
}
\quad
\subfigure{
\includegraphics[width=8.5cm]{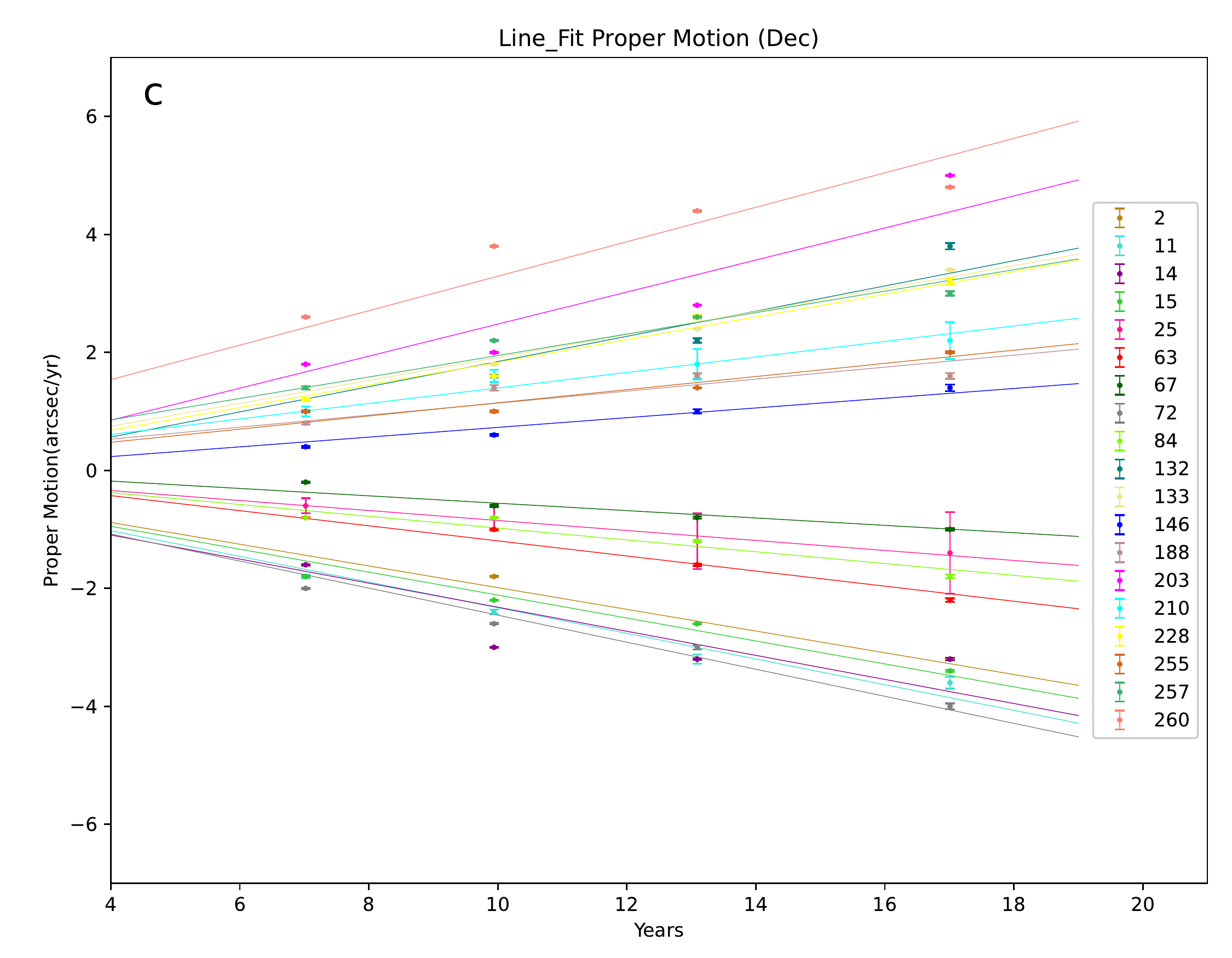}
}
\quad
\subfigure{
\includegraphics[width=8.5cm]{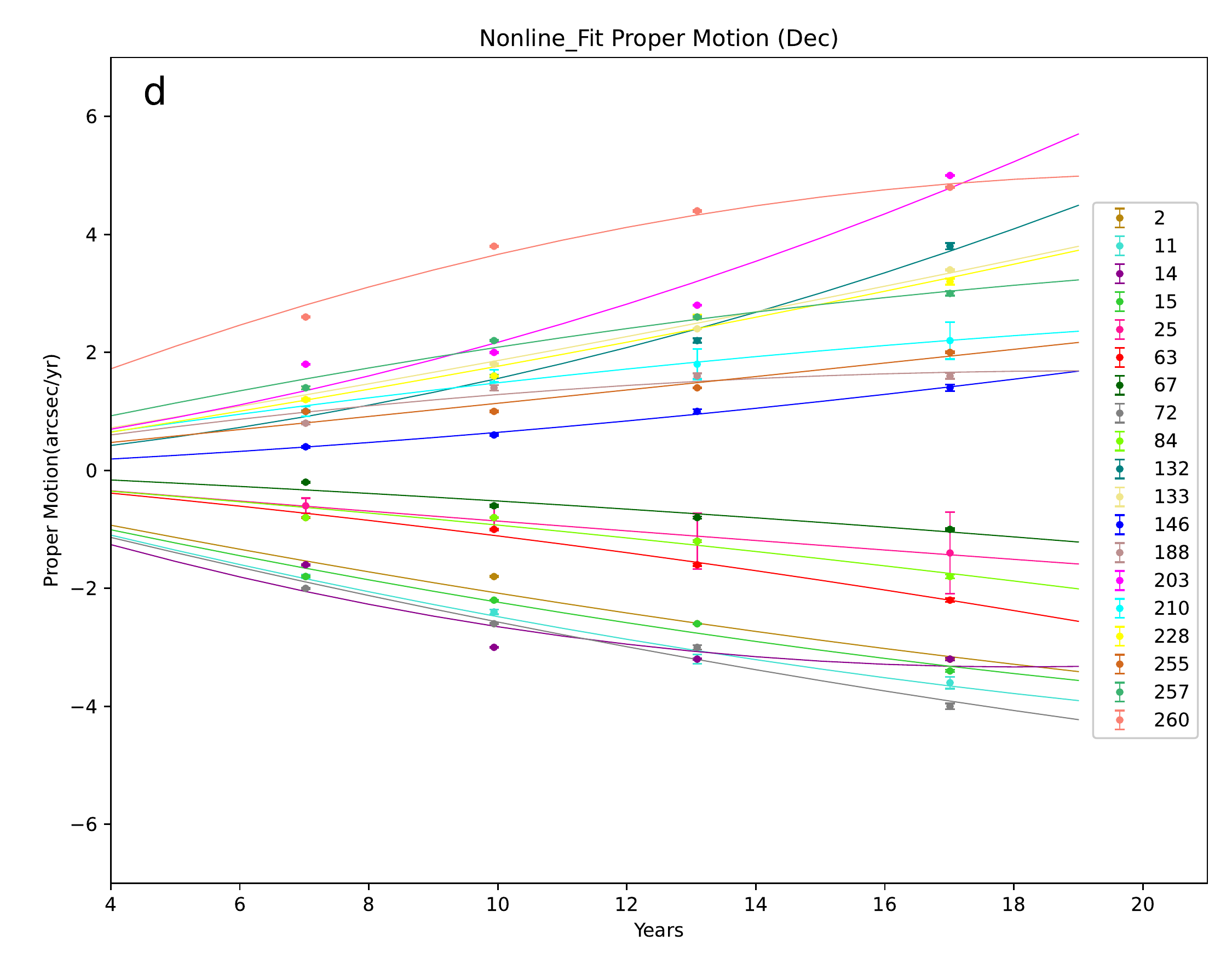}
}
\caption{Fitting of linear and nonlinear models to partial knots' motion in Cas A. The figure "a" and "c" are the knots' proper motion fitted by linear model in Ra and Dec, respectively. The figure "b" and "d" are the knots' proper motion fitted by nonlinear model in Ra and Dec, respectively.}\label{knot_fits}
\end{figure}

\begin{figure}
\centering
\includegraphics[width=9.8cm]{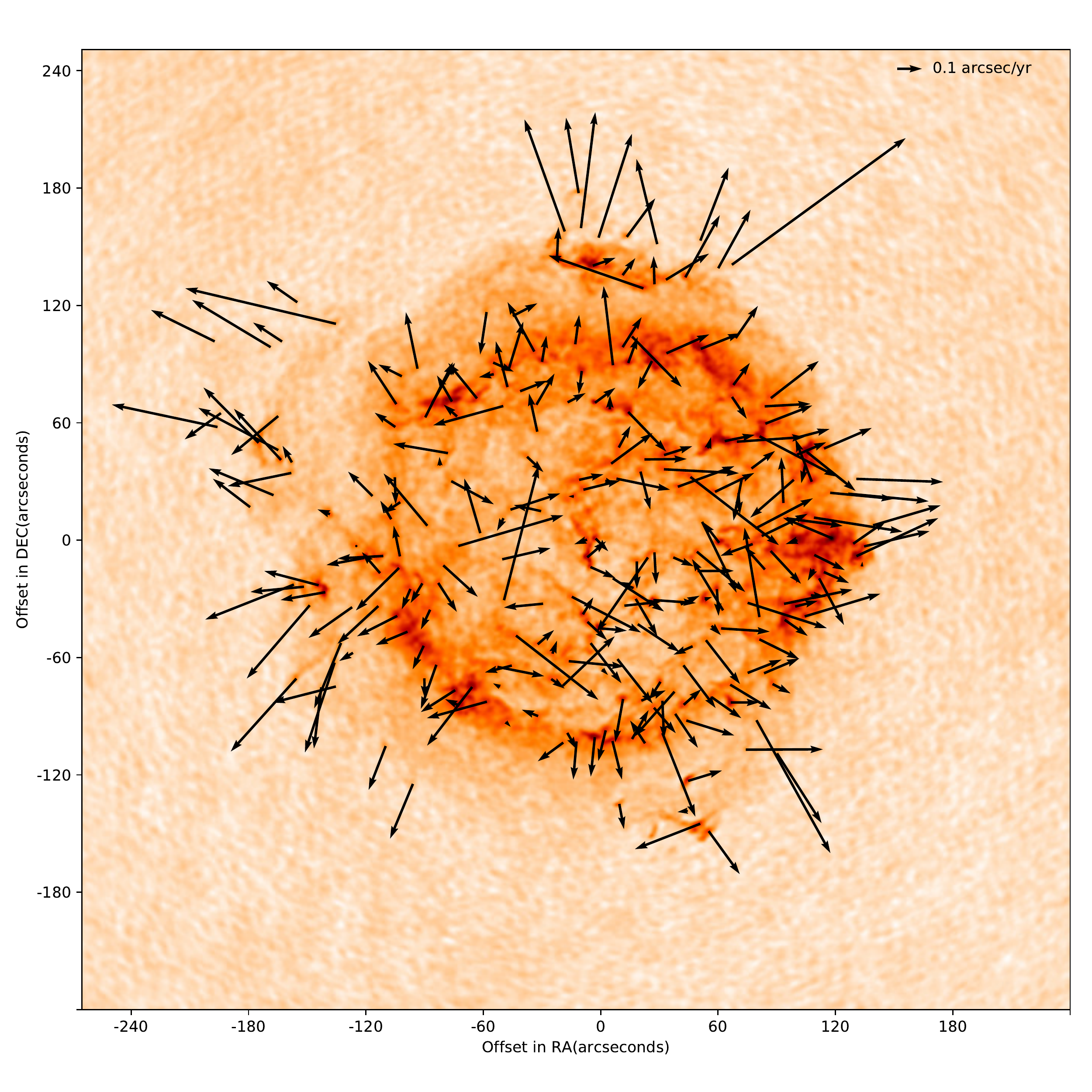}
\caption{The spatial distribution of proper motion of 260 radio knots.}\label{proper_motion}
\end{figure}

We have used nonlinear fitting method to calculate the positional shifts ($\Delta\alpha$ and $\Delta\sigma$) as a quadratic function of evolution time (i.e., $\Delta{\rm s}=v_{0}t+a t^2$, $\Delta{\rm s}$ is the positional shifts of right ascension or declination), and obtained the proper motion of each radio features. The date of each VLA images adopted in the fitting is the average date of the A-, B-, C- and D-configuration observations of VLA. Most radio emissions in Cas A have decelerated significantly, nonlinear fitting can describe the real evolution of its motion better. From the linear and nonlinear model fits of partial knots' motion shown in Figure \ref{knot_fits}, linear fitting model does not provide a good fit to knots' motion trajectories, especially for knots with fast motion. The spatial distribution of proper motion vectors of the 260 radio knots is shown in Figure \ref{proper_motion}. Comparing with the plot of proper motion vectors in AR95, there is no significant systematic differences.

\begin{figure}[b!]
\centering
\subfigure{
\includegraphics[width=8.5cm]{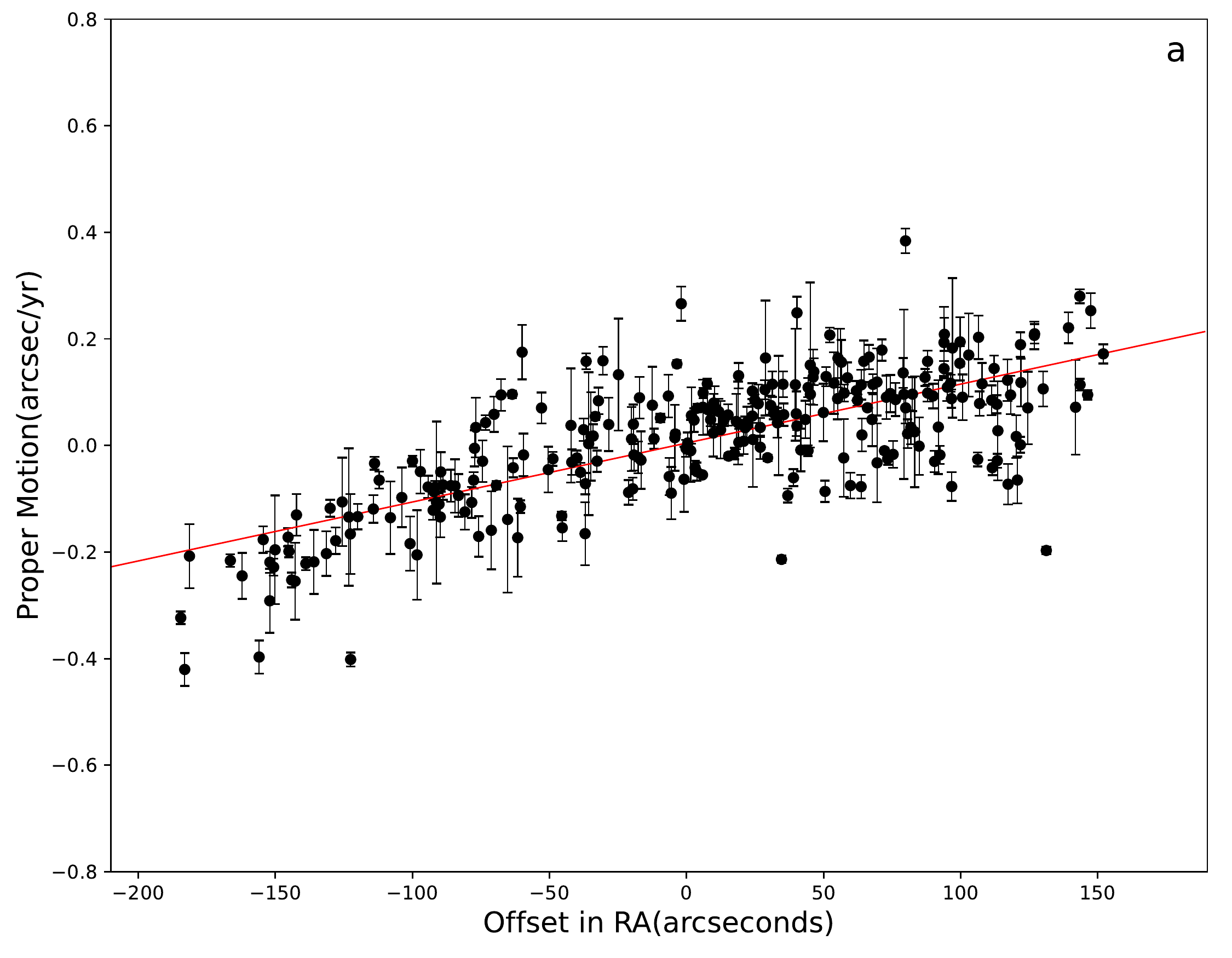}
}
\quad
\subfigure{
\includegraphics[width=8.5cm]{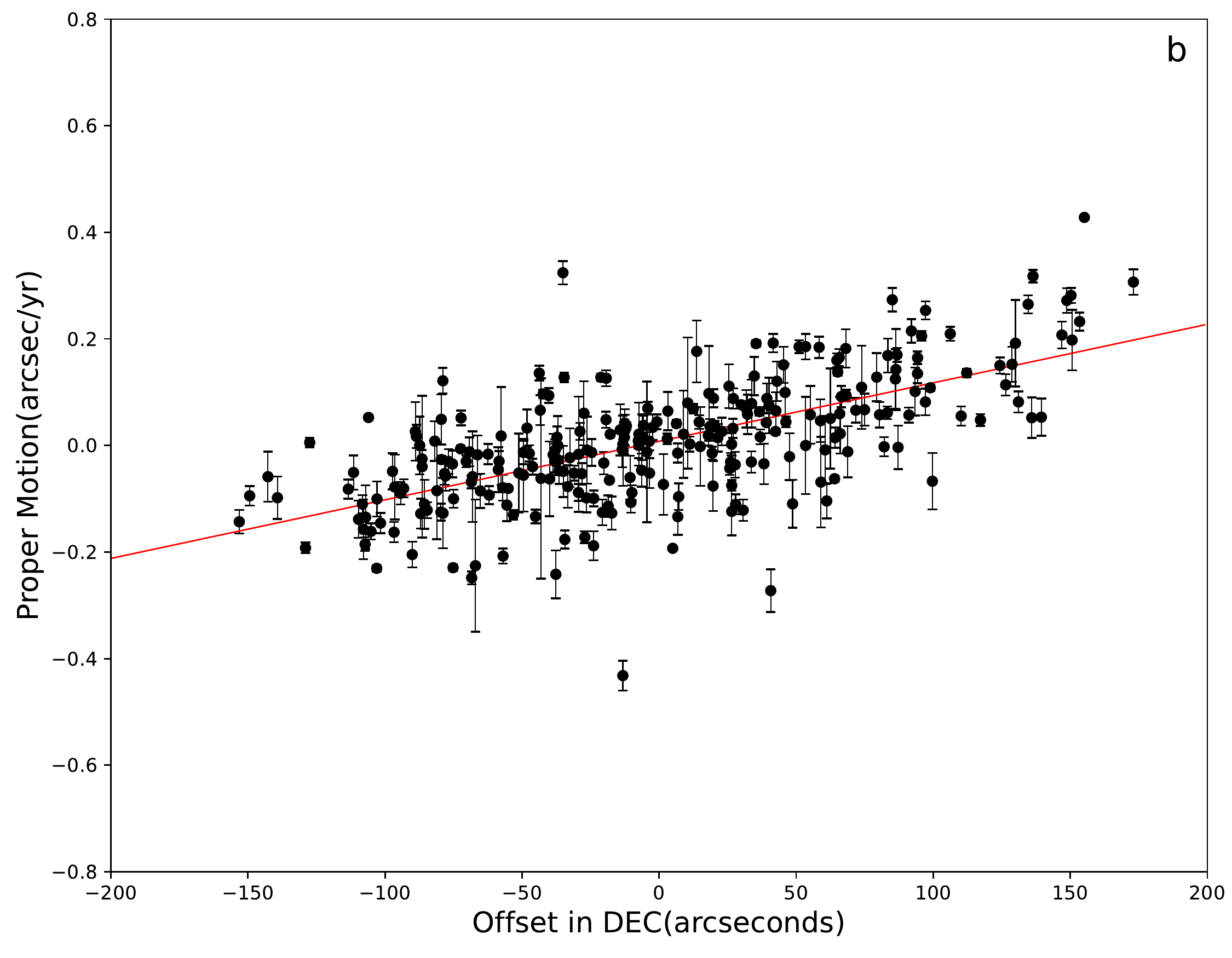}
}
\caption{a: Linear weighted fits to the right ascension of proper motion as a function of radio knot position; b: Linear weighted fits to the declination of proper motion as a function of radio knot position}\label{expasion_age}
\end{figure}

The location of radio expansion center is at $\alpha(1950)=23^{\rm h}21^{\rm m}9^{\rm s}_{\cdot}7 \pm 0^{\rm s}_{\cdot}29$, $\delta(1950)=+58^{\circ}32^{\prime}25^{\prime\prime}_{\cdot}2 \pm 2^{\prime\prime}_{\cdot}2$ from our measured proper motion results, which is consistent with the center position measured in AR95 within the quoted errors. A linear relationship between proper motion and distance from the radio expansion center has been used to deduce bulk expansion timescales of the remnant in AR95 and \citet{1986MNRAS.219...13T}, which assumes that all radio emission have same initial velocity and does not show deceleration or experience the same deceleration process.

To track the motion evolution of Cas A, we used the same linear expansion model to estimate the expansion timescales of the remnant in right ascension and declination. Before fitting the relationship between proper motion and distance from the expansion center, we used the linear fitting method to recalculated the proper motion of radio features to ensure the rationality of this method. Figure \ref{expasion_age} shows the fitting results of expansion timescales, the expansion timescales of right ascension is ${\rm T}_{\rm RA}=904 \pm 10$ yrs, which equivalent to the expansion timescale of declination ${\rm T}_{\rm DEC}=910 \pm 12$ yrs. Our measured expansion timescales is conflicts with \citeauthor{1986MNRAS.219...13T}' (\citeyear{1986MNRAS.219...13T}) measurement that the expansion timescales of right ascension was longer than declination significantly. Intriguingly, the expansion timescale in right ascension is approximates the expansion timescale in declination, which is consistent with AR95 (${\rm T}_{\rm RA}=866 \pm 8$ yrs, ${\rm T}_{\rm DEC}=861 \pm 9$ yrs). However, the expansion timescales we obtained are longer. Under the assumption of a same expansion model we detected longer expansion timescales, which indicates that radio emission in the remnant were indeed deceleration.

The knots located at the northeastern and northern edges have relative high proper motion from the Figure \ref{proper_motion}, the velocity can reach to $\sim$14500 km $\rm s^{-1}$ in the north. As shown in Figure \ref{expasion_age}a, the knots in located at the east and west under the fitting line are not well fitted, which indicates that knots have lower proper motion at the western and higher proper motion at the eastern. In the Figure \ref{expasion_age}b, the knots located at the northern are above the fitting line, indicating that the knots with higher proper motion in the north.

\begin{figure}[t!]
\centering
\includegraphics[width=8.8cm]{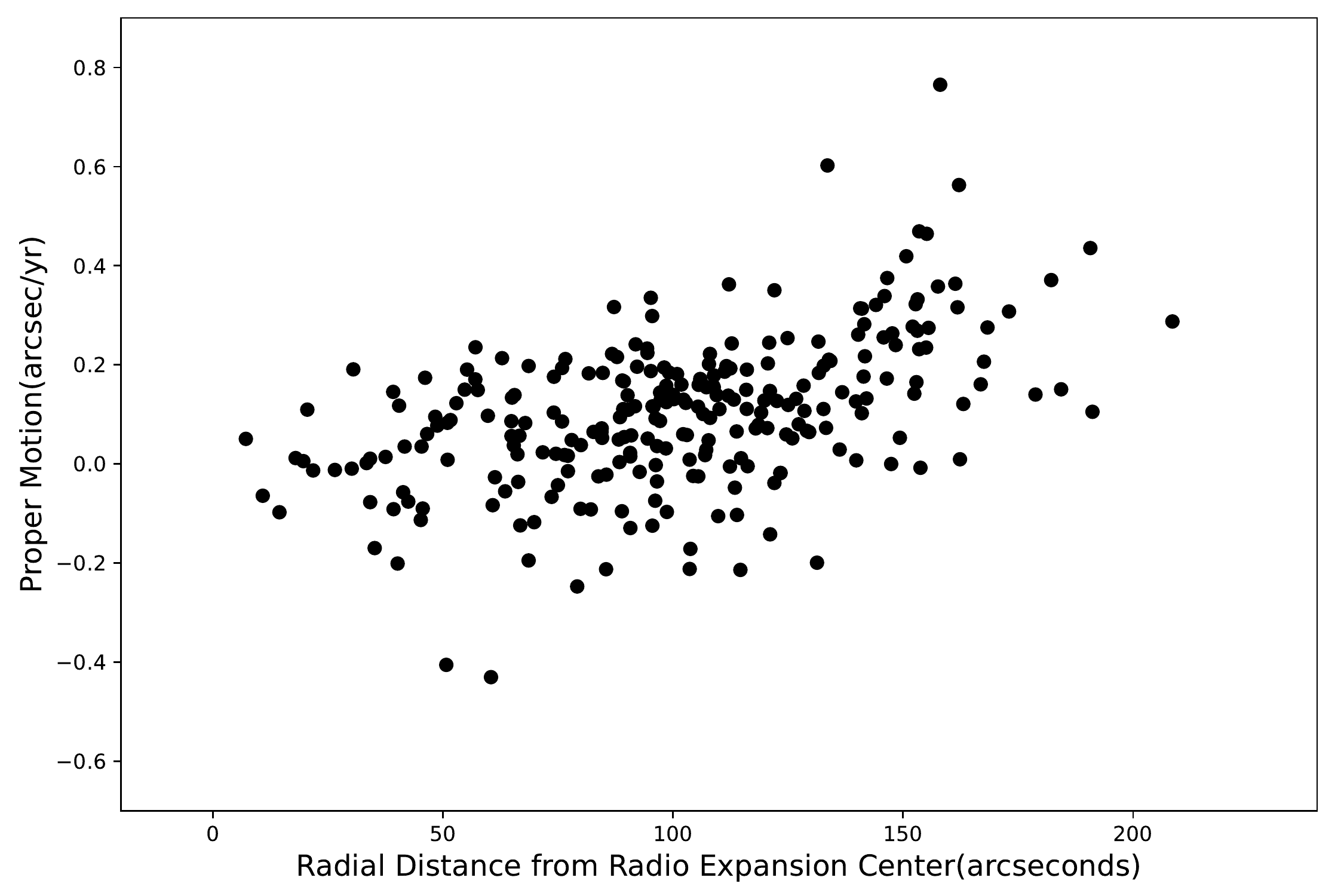}
\caption{Projected radial proper motion vs. the distance from the radio expansion center.}\label{radial_ProperMotion}
\end{figure}

Ejecta interacting with the reverse shock forms a bright radio emission circle with a radius of $\sim$1.7 pc in Cas A, and a fainter plateau extending to a radius of $\sim$ 2.5 pc (\citealp{2014ApJ...785..130Z}; \citealp{2014ApJ...785....7D}). This radio morphology corresponds to the two knots groups distributed in Figure \ref{radial_ProperMotion}, which can be understood as concentrically expanding shells as in AR95. Knots experience varying degrees of deceleration at different shells, and the proper motion of knots at the outer shell are faster than at the inner shell obviously. Moreover, the radial expansion ratio of radio emission features outside the bright radio ring increases with the distance from the radio expansion center. The average rate of expansion about 2100 km $\rm s^{-1}$, while the nonlinearity values outer the bright radio ring is $\sim$2550 km $\rm s^{-1}$, and $\sim$1750 km $\rm s^{-1}$ inner the bright radio ring. This expansion timescales in the bright ring region is $\sim$857 yrs, and the outer region is $\sim$525 yrs. The outer radio emission dynamic age of our measured close to the overall expansion timescales was measured by \citet{1999MNRAS.305..957A} (i.e., $\sim$400-500 yrs). Interaction between the remnant and molecular clouds hints a deceleration of knots in the shell (\citealp{2014ApJ...796..144K}; \citealp{2018ApJ...853...46S}), outer shorter dynamics timescales indicate a smaller deceleration of knots located at the outer shell.

\begin{figure}[b]
\centering
\includegraphics[width=8.8cm]{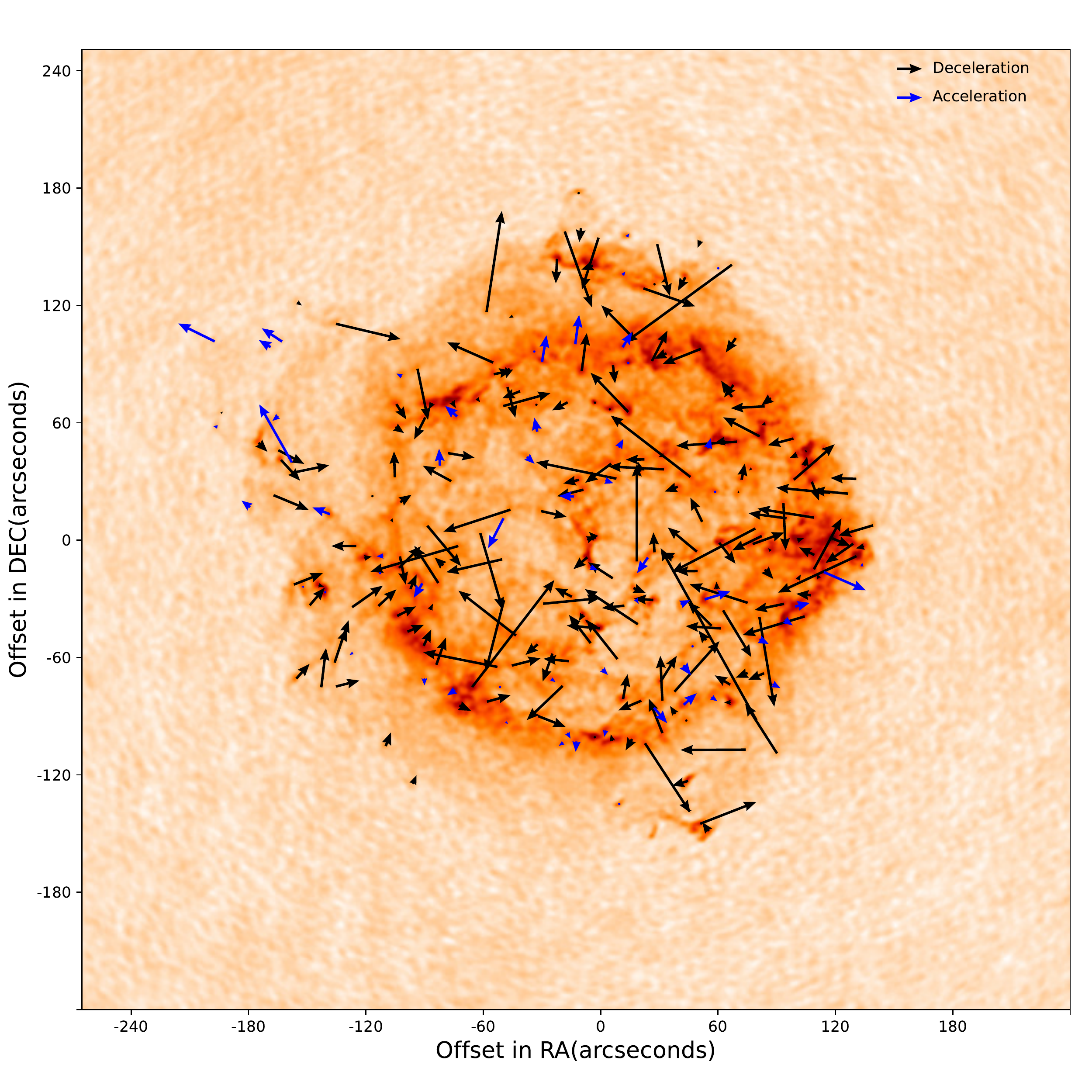}
\caption{The spatial distribution of the acceleration vector projected in the proper motion vectors of the 260 radio knots. Black: the decelerated radio knots projected on the proper motion vector; Blue: the accelerated radio knots.}\label{projected_acceleration}
\end{figure}

\begin{figure*}[t!]
\centering
\includegraphics[width=14cm]{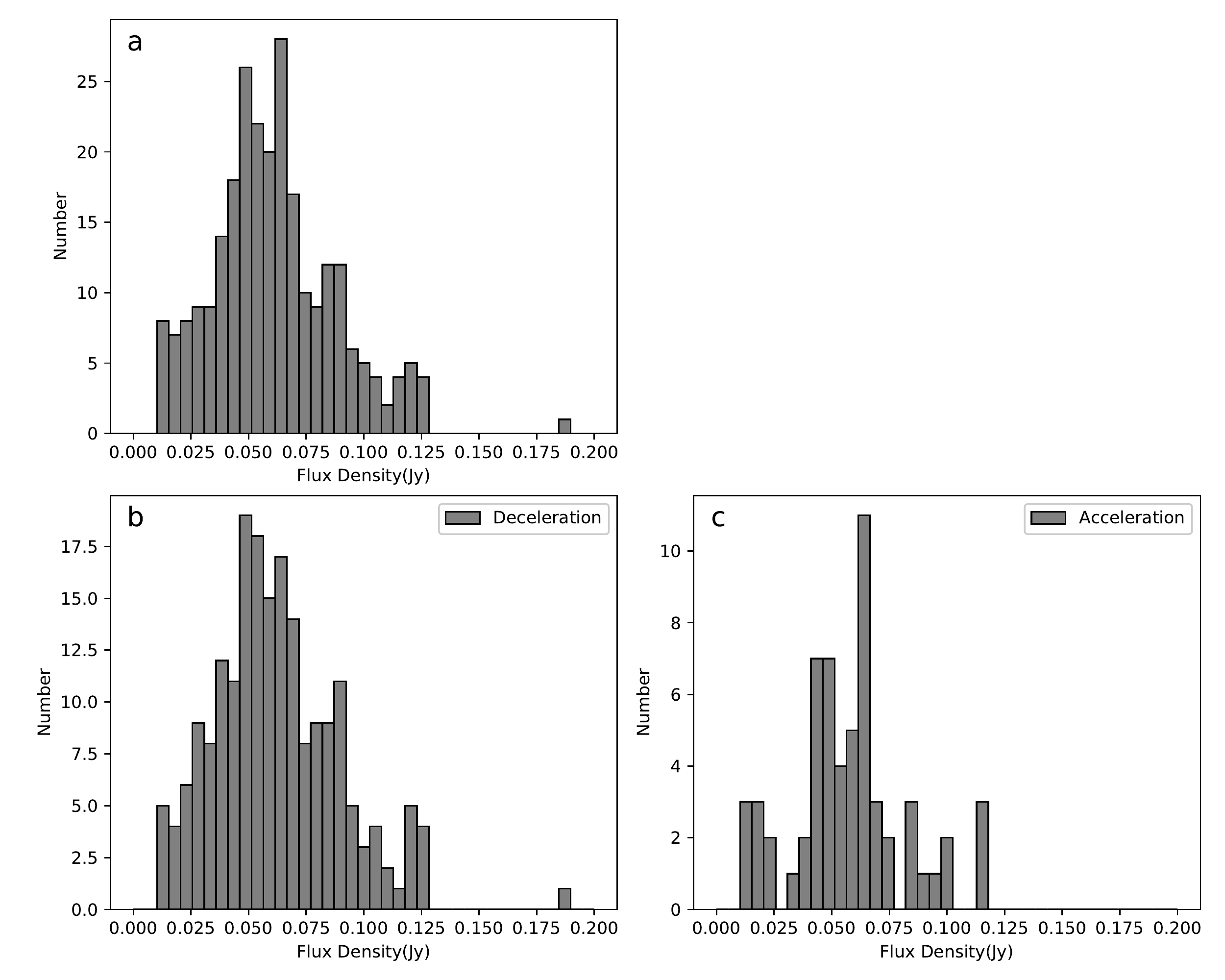}
\caption{The histogram of peak flux of compact features we studied.}\label{peak_flux_distribution}
\end{figure*}

\subsection{Accelerations of Compact Radio Knots}
Longer expansion timescales have been detected using a same linear expansion model as for AR95, which indicates that the knots in Cas A have already decelerating significantly. In Section \ref{motion_result}, the proper motion of 260 knots have been measured by using nonlinear fitting model ($\Delta{\rm s}=v_{0}t+a_{\rm s} \rm t^2$) to fit position shifts ($\Delta\alpha$,$ \Delta\sigma$), meanwhile the magnitude of accelerations $\rm a_{\rm s}$ has been given. For part knots, the computations of the position shifts appear in the opposite direction for some epochs, and the data of these epochs were excluded in fitting. For another, in order to calculate the accelerations of each knot, at least four epoch data are kept in the fitting (including the reference image). Therefore, part of the proper motions and the accelerations are calculated only based on data from four epochs. Those knots that position shifts in the opposite direction more than two epochs has been excluded in determining the used samples.

We have recalculated the accelerations of each knot by sequentially excluding each epoch from the fitting, the average significant level of $a_{\rm RA}$ and $a_{\rm DEC}$ around 2$\sigma$. For some knots, the obtained accelerations are less than the associated errors, which may be cause by the changes in shape or confusion with other emissions. Checking the spatial distribution of the accelerations, the accelerations field is somewhat disordered but not random noise. To determine individual knots is accelerating or decelerating directly, we project the accelerations onto the direction of its proper motion vectors. The spatial distribution of the acceleration vectors projection is displayed in Figure \ref{projected_acceleration}, which shows that $\sim$76\% of knots are decelerating. Taking as a whole, the components of the accelerations in the radial coordinates, right ascension and declination are $\rm a_{\rm radial}=-3.9\pm2.0$ mas $\rm yr^{-2}$, $\rm a_{\rm RA}=-1.7\pm0.7$ mas $\rm yr^{-2}$, $\rm a_{\rm DEC}=-2.6\pm1.0$ mas $\rm yr^{-2}$, respectively.

\citet{2022ApJ...929...57V} reported the proper motion of forward-  and reverse-shock of Cas A in X-ray, and demonstrated that the forward shock in the western is accelerating. Most of the knots located at the outer edge of the northeast are shown accelerating from the measurement result displayed in Figure \ref{projected_acceleration}, but the spatial distribution of acceleration vectors in other regions appears to have no obvious regulations. The histogram peak flux in Figure \ref{peak_flux_distribution} shows that there is no obvious difference between the acceleration and deceleration distribution, which hints that the acceleration or deceleration of individual knot have not correlation with the flux density distribution.

For the acceleration of radio emissions, we think there are two reasons. One is that the accelerated knots confuses with other radio emissions. Knots which locate in bright emission regions are more easily confuse with other radio emissions, and the acceleration is generally less than the quoted error. The second is that the stellar wind density of the remnant's progenitor is spatially inhomogeneous. Similarity in the morphology and spectra of the adjacent nebulae with the eastern cloud of Cas A indicates the progenitor's mass-loss \citep{2020ApJ...891..116W}. Mass-loss leads to the fluctuations in stellar wind density of progenitor, which will accelerate while a knot moves from high density region to low density region.

\subsection{Brightness Changes of Compact Radio Knots}
\begin{figure}[t]
\centering
\includegraphics[width=8.8cm]{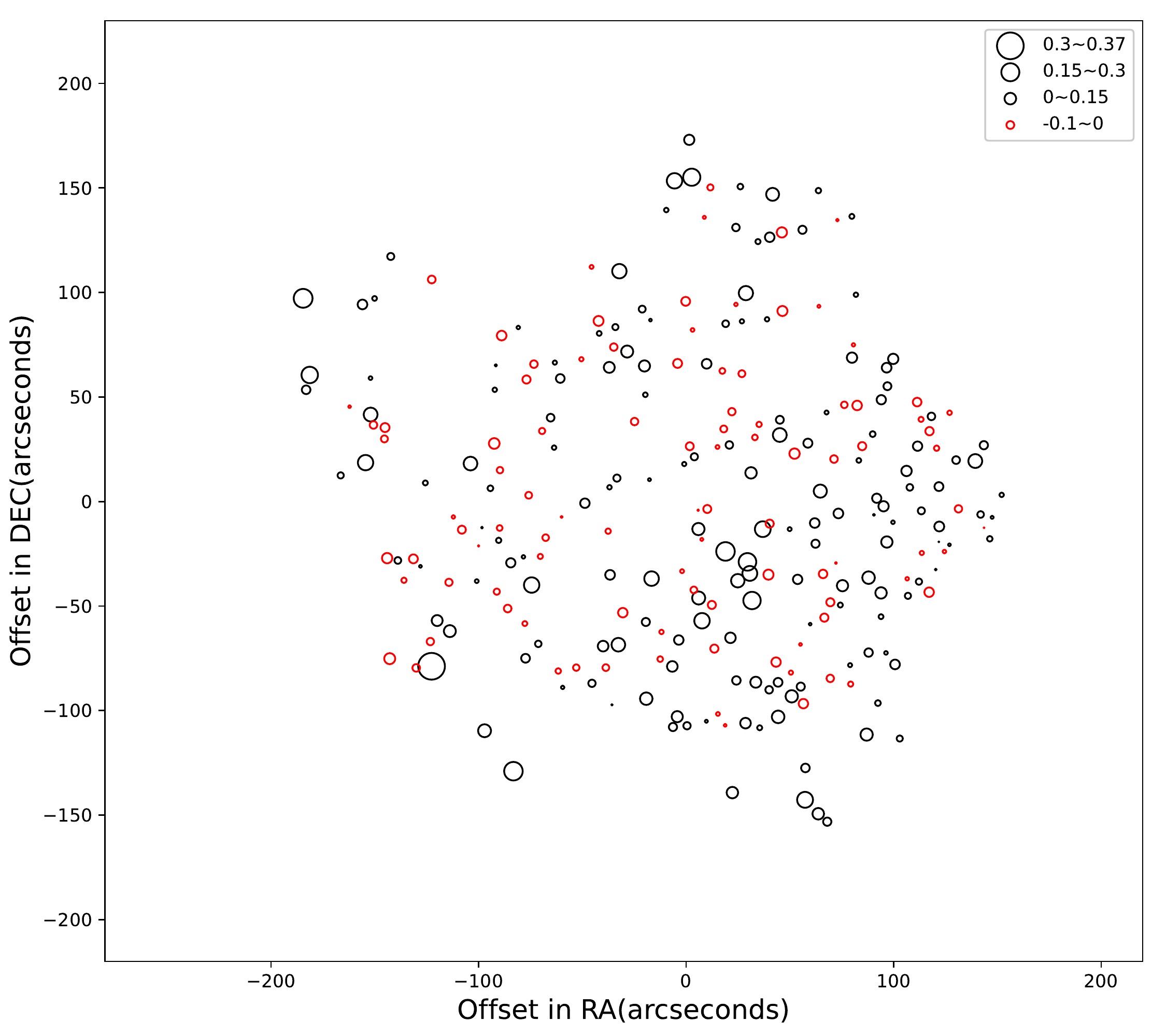}
\caption{The fractional spatial distribution of brightness images of 260 compact radio knots.}\label{brightness_distribution}
\end{figure}

In the brightness change measurement, to avoid the influence of the difference of synthetic aperture coverage in difference epochs on the flux density and the shape of large-scale structures, we have smoothed all continuum images into a same beamsize (i.e., $1.5^{\prime\prime} \times 1.5^{\prime\prime}$). Brightness changes have been measured using the least square method in Section \ref{measurement_position}, and the brightness change fraction was derived by nonlinear fitting. The spatial distribution of fractional changes in brightness is displayed in Figure \ref{brightness_distribution}. The flux of Cas A is constantly fading (\citealp{2000ApJ...537..904R}; \citealp{2017MNRAS.469.1299T}). However, our measurement shows that $\sim$62\% of the radio knots are brightening with an average rate of $\sim$2.2\%, these knots occupy only a small fraction of the total emission of Cas A.

AR95 has reported that the large-scale emission and compact features in Cas A are brightening with an average rate of $\sim$1.6\% $\rm yr^{-1}$. In the current work, $\sim$62\% of the compact features are brightening, and the brightness changes fraction varies in different regions. The compact features on the bright radio ring ($90^{\prime\prime}$-$130^{\prime\prime}$) have a relatively low brightening rate of $\sim$1.4\% $\rm yr^{-1}$. The average brightening rate of compact features in the inner of the bright radio ring ($\leq 90^{\prime\prime}$) is $\sim$2.1\% $\rm yr^{-1}$, this value is mainly the contribution of the southwest region inside the bright radio ring, which is brightening rapidly. In the diffuse plateau region, compact features are brightening at a relatively fast rate of 3.7\% $\rm yr^{-1}$.

Bright radio emissions rely on the synchrotron radiation emitted by electrons moving in magnetic field, a high radio emission means a high magnetic field for a given number density of electrons. Most of the radio-bright knots are distributed in the bright ring of Cas A from Figure \ref{knots_distribution}, whereas the primary site of cosmic-ray acceleration located in the forward shock front (\citealp{2003ApJ...584..758V}; \citealp{2012SSRv..173..369H}). Moreover, studies of radio spectral index have shown no evidence of reacceleration in bright radio knots (\citealp{1996ApJ...456..234A}; \citealp{1999ApJ...518..284W}). These studies indicate that brightness variations are links to the fluctuations of magnetic field directly. Therefore, the fading, brightening, disappearing or new appearing of compact features in Cas A are likely to be the result of magnetic field variations.

\section{summary}
We have obtained five images of Cas A at 5 GHz in all four configurations of the VLA from 1987 to 2004, while all images have been smoothed to the same resolution of $1.5^{\prime\prime}$. We measured the proper motion and brightness changes of compact radio features in Cas A, gave the radio expansion center locates at  $\alpha(1950)=23^{\rm h}21^{\rm m}9^{\rm s}_{\cdot}7 \pm 0^{\rm s}_{\cdot}29$, $\delta(1950)=+58^{\circ}32^{\prime}25^{\prime\prime}_{\cdot}2 \pm 2^{\prime\prime}_{\cdot}$, and the expansion timescales of right ascension and declination: $T_{\rm RA}=904 \pm 10$ yrs, $T_{\rm DEC}=910 \pm 12$ yrs, respectively. Because of the limit of angular resolution of observation, the uncertainty of proper motion measurement will lead to different estimation for the expansion timescale. Benefiting from the improvement in angular resolution, the expansion timescale is more reliable.

The remnant is overall in a decelerating phase, but only about three-quarters of compact radio knots are decelerating. Some radio knots with acceleration are in the region of radio emission confusion and these radio knots may not really be in acceleration. In addition to the effect of measurement errors in the radio emission confusion regions, the compact features' acceleration phenomenon suggests that there is density fluctuation in the stellar wind of Cas A's progenitor. During 17 yrs, the brightness of the compact radio features in Cas A has changed significantly, which suggests that the magnetic field is changing rapidly.

\begin{acknowledgments}
We thank Dr. Shi Hui and Dr. Yang Ai-Yuan for the guidance on data calibration in this work. We aknowledge support from the National Key R\&D Programs of China (2018YFA0404203, 2018YFA0404202), National Natural Science Foundation of China (12041301, 12073039) and the China Manned Space Project (CMS-CSST-2021-A09) . We thank the staff of the  Karl G.Jansky Very Larege Array for the observation of Cassiopeia A.
\end{acknowledgments}

\bibliography{Cassiopeia_A}{}

\begin{thebibliography}{}
\expandafter\ifx\csname natexlab\endcsname\relax\def\natexlab#1{#1}\fi
\providecommand{\url}[1]{\href{#1}{#1}}
\providecommand{\dodoi}[1]{doi:~\href{http://doi.org/#1}{\nolinkurl{#1}}}
\providecommand{\doeprint}[1]{\href{http://ascl.net/#1}{\nolinkurl{http://ascl.net/#1}}}
\providecommand{\doarXiv}[1]{\href{https://arxiv.org/abs/#1}{\nolinkurl{https://arxiv.org/abs/#1}}}

\bibitem[{{Ag{\"u}eros} \& {Green}(1999)}]{1999MNRAS.305..957A}
{Ag{\"u}eros}, M.~A., \& {Green}, D.~A. 1999, \mnras, 305, 957,
  \dodoi{10.1046/j.1365-8711.1999.02504.x}

\bibitem[{{Ahnen} {et~al.}(2017){Ahnen}, {Ansoldi}, {Antonelli}, {Arcaro},
  {Babi{\'c}}, {Banerjee}, {Bangale}, {Barres de Almeida}, {Barrio}, {Becerra
  Gonz{\'a}lez}, {Bednarek}, {Bernardini}, {Berti}, {Bhattacharyya},
  {Biasuzzi}, {Biland}, {Blanch}, {Bonnefoy}, {Bonnoli}, {Carosi}, {Carosi},
  {Chatterjee}, {Colak}, {Colin}, {Colombo}, {Contreras}, {Cortina}, {Covino},
  {Cumani}, {Da Vela}, {Dazzi}, {De Angelis}, {De Lotto}, {de O{\~n}a
  Wilhelmi}, {Di Pierro}, {Doert}, {Dom{\'\i}nguez}, {Dominis Prester},
  {Dorner}, {Doro}, {Einecke}, {Eisenacher Glawion}, {Elsaesser},
  {Engelkemeier}, {Fallah Ramazani}, {Fern{\'a}ndez-Barral}, {Fidalgo},
  {Fonseca}, {Font}, {Fruck}, {Galindo}, {Garc{\'\i}a L{\'o}pez},
  {Garczarczyk}, {Gaug}, {Giammaria}, {Godinovi{\'c}}, {Gora}, {Guberman},
  {Hadasch}, {Hahn}, {Hassan}, {Hayashida}, {Herrera}, {Hose}, {Hrupec},
  {Inada}, {Ishio}, {Konno}, {Kubo}, {Kushida}, {Kuve{\v{z}}di{\'c}}, {Lelas},
  {Lindfors}, {Lombardi}, {Longo}, {L{\'o}pez}, {Maggio}, {Majumdar},
  {Makariev}, {Maneva}, {Manganaro}, {Mannheim}, {Maraschi}, {Mariotti},
  {Mart{\'\i}nez}, {Mazin}, {Menzel}, {Minev}, {Mirzoyan}, {Moralejo},
  {Moreno}, {Moretti}, {Neustroev}, {Niedzwiecki}, {Nievas Rosillo}, {Nilsson},
  {Ninci}, {Nishijima}, {Noda}, {Nogu{\'e}s}, {Paiano}, {Palacio}, {Paneque},
  {Paoletti}, {Paredes}, {Pedaletti}, {Peresano}, {Perri}, {Persic}, {Prada
  Moroni}, {Prandini}, {Puljak}, {Garcia}, {Reichardt}, {Rhode}, {Rib{\'o}},
  {Rico}, {Righi}, {Saito}, {Satalecka}, {Schroeder}, {Schweizer}, {Shore},
  {Sitarek}, {{\v{S}}nidari{\'c}}, {Sobczynska}, {Stamerra}, {Strzys},
  {Suri{\'c}}, {Takalo}, {Tavecchio}, {Temnikov}, {Terzi{\'c}}, {Tescaro},
  {Teshima}, {Torres-Alb{\`a}}, {Treves}, {Vanzo}, {Vazquez Acosta}, {Vovk},
  {Ward}, {Will}, \& {Zari{\'c}}}]{2017MNRAS.472.2956A}
{Ahnen}, M.~L., {Ansoldi}, S., {Antonelli}, L.~A., {et~al.} 2017, \mnras, 472,
  2956, \dodoi{10.1093/mnras/stx2079}

\bibitem[{{Allen} {et~al.}(1997){Allen}, {Keohane}, {Gotthelf}, {Petre},
  {Jahoda}, {Rothschild}, {Lingenfelter}, {Heindl}, {Marsden}, {Gruber},
  {Pelling}, \& {Blanco}}]{1997ApJ...487L..97A}
{Allen}, G.~E., {Keohane}, J.~W., {Gotthelf}, E.~V., {et~al.} 1997, \apjl, 487,
  L97, \dodoi{10.1086/310878}

\bibitem[{{Anderson} \& {Rudnick}(1995)}]{1995ApJ...441..307A}
{Anderson}, M.~C., \& {Rudnick}, L. 1995, \apj, 441, 307,
  \dodoi{10.1086/175357}

\bibitem[{{Anderson} \& {Rudnick}(1996)}]{1996ApJ...456..234A}
---. 1996, \apj, 456, 234, \dodoi{10.1086/176644}

\bibitem[{{Ashworth}(1980)}]{1980JHA....11....1A}
{Ashworth}, W.~B., J. 1980, Journal for the History of Astronomy, 11, 1,
  \dodoi{10.1177/002182868001100102}

\bibitem[{{Bell}(1977)}]{1977MNRAS.179..573B}
{Bell}, A.~R. 1977, \mnras, 179, 573, \dodoi{10.1093/mnras/179.4.573}

\bibitem[{{Bell} {et~al.}(1975){Bell}, {Gull}, \&
  {Kenderdine}}]{1975Natur.257..463B}
{Bell}, A.~R., {Gull}, S.~F., \& {Kenderdine}, S. 1975, \nat, 257, 463,
  \dodoi{10.1038/257463a0}

\bibitem[{{Chevalier} \& {Kirshner}(1978)}]{1978ApJ...219..931C}
{Chevalier}, R.~A., \& {Kirshner}, R.~P. 1978, \apj, 219, 931,
  \dodoi{10.1086/155855}

\bibitem[{{Chevalier} \& {Kirshner}(1979)}]{1979ApJ...233..154C}
---. 1979, \apj, 233, 154, \dodoi{10.1086/157377}

\bibitem[{{Cornwell}(2008)}]{2008ISTSP...2..793C}
{Cornwell}, T.~J. 2008, IEEE Journal of Selected Topics in Signal Processing,
  2, 793, \dodoi{10.1109/JSTSP.2008.2006388}

\bibitem[{{DeLaney} {et~al.}(2014){DeLaney}, {Kassim}, {Rudnick}, \&
  {Perley}}]{2014ApJ...785....7D}
{DeLaney}, T., {Kassim}, N.~E., {Rudnick}, L., \& {Perley}, R.~A. 2014, \apj,
  785, 7, \dodoi{10.1088/0004-637X/785/1/7}

\bibitem[{{DeLaney} \& {Rudnick}(2003)}]{2003ApJ...589..818D}
{DeLaney}, T., \& {Rudnick}, L. 2003, \apj, 589, 818, \dodoi{10.1086/374813}

\bibitem[{{DeLaney} {et~al.}(2004){DeLaney}, {Rudnick}, {Fesen}, {Jones},
  {Petre}, \& {Morse}}]{2004ApJ...613..343D}
{DeLaney}, T., {Rudnick}, L., {Fesen}, R.~A., {et~al.} 2004, \apj, 613, 343,
  \dodoi{10.1086/422906}

\bibitem[{{Dickel} \& {Greisen}(1979)}]{1979A&A....75...44D}
{Dickel}, J.~R., \& {Greisen}, E.~W. 1979, \aap, 75, 44

\bibitem[{{Fesen}(2001)}]{2001ApJS..133..161F}
{Fesen}, R.~A. 2001, \apjs, 133, 161, \dodoi{10.1086/319181}

\bibitem[{{Fesen} {et~al.}(1987){Fesen}, {Becker}, \&
  {Blair}}]{1987ApJ...313..378F}
{Fesen}, R.~A., {Becker}, R.~H., \& {Blair}, W.~P. 1987, \apj, 313, 378,
  \dodoi{10.1086/164974}

\bibitem[{{Fesen} {et~al.}(1988){Fesen}, {Becker}, \&
  {Goodrich}}]{1988ApJ...329L..89F}
{Fesen}, R.~A., {Becker}, R.~H., \& {Goodrich}, R.~W. 1988, \apjl, 329, L89,
  \dodoi{10.1086/185183}

\bibitem[{{Fesen} {et~al.}(2001){Fesen}, {Morse}, {Chevalier}, {Borkowski},
  {Gerardy}, {Lawrence}, \& {van den Bergh}}]{2001AJ....122.2644F}
{Fesen}, R.~A., {Morse}, J.~A., {Chevalier}, R.~A., {et~al.} 2001, \aj, 122,
  2644, \dodoi{10.1086/323539}

\bibitem[{{Fesen} {et~al.}(2006){Fesen}, {Hammell}, {Morse}, {Chevalier},
  {Borkowski}, {Dopita}, {Gerardy}, {Lawrence}, {Raymond}, \& {van den
  Bergh}}]{2006ApJ...645..283F}
{Fesen}, R.~A., {Hammell}, M.~C., {Morse}, J., {et~al.} 2006, \apj, 645, 283,
  \dodoi{10.1086/504254}

\bibitem[{{Gotthelf} {et~al.}(2001){Gotthelf}, {Koralesky}, {Rudnick}, {Jones},
  {Hwang}, \& {Petre}}]{2001ApJ...552L..39G}
{Gotthelf}, E.~V., {Koralesky}, B., {Rudnick}, L., {et~al.} 2001, \apjl, 552,
  L39, \dodoi{10.1086/320250}

\bibitem[{{Grefenstette} {et~al.}(2015){Grefenstette}, {Reynolds}, {Harrison},
  {Humensky}, {Boggs}, {Fryer}, {DeLaney}, {Madsen}, {Miyasaka}, {Wik},
  {Zoglauer}, {Forster}, {Kitaguchi}, {Lopez}, {Nynka}, {Christensen}, {Craig},
  {Hailey}, {Stern}, \& {Zhang}}]{2015ApJ...802...15G}
{Grefenstette}, B.~W., {Reynolds}, S.~P., {Harrison}, F.~A., {et~al.} 2015,
  \apj, 802, 15, \dodoi{10.1088/0004-637X/802/1/15}

\bibitem[{{Helder} \& {Vink}(2008)}]{2008ApJ...686.1094H}
{Helder}, E.~A., \& {Vink}, J. 2008, \apj, 686, 1094, \dodoi{10.1086/591242}

\bibitem[{{Helder} {et~al.}(2012){Helder}, {Vink}, {Bykov}, {Ohira}, {Raymond},
  \& {Terrier}}]{2012SSRv..173..369H}
{Helder}, E.~A., {Vink}, J., {Bykov}, A.~M., {et~al.} 2012, \ssr, 173, 369,
  \dodoi{10.1007/s11214-012-9919-8}

\bibitem[{{Hogg} {et~al.}(1969){Hogg}, {MacDonald}, {Conway}, \&
  {Wade}}]{1969AJ.....74.1206H}
{Hogg}, D.~E., {MacDonald}, G.~H., {Conway}, R.~G., \& {Wade}, C.~M. 1969, \aj,
  74, 1206, \dodoi{10.1086/110924}

\bibitem[{{Hughes} {et~al.}(2000){Hughes}, {Rakowski}, {Burrows}, \&
  {Slane}}]{2000ApJ...528L.109H}
{Hughes}, J.~P., {Rakowski}, C.~E., {Burrows}, D.~N., \& {Slane}, P.~O. 2000,
  \apjl, 528, L109, \dodoi{10.1086/312438}

\bibitem[{{Hwang} \& {Laming}(2012)}]{2012ApJ...746..130H}
{Hwang}, U., \& {Laming}, J.~M. 2012, \apj, 746, 130,
  \dodoi{10.1088/0004-637X/746/2/130}

\bibitem[{{Inoue} {et~al.}(2012){Inoue}, {Yamazaki}, {Inutsuka}, \&
  {Fukui}}]{2012ApJ...744...71I}
{Inoue}, T., {Yamazaki}, R., {Inutsuka}, S.-i., \& {Fukui}, Y. 2012, \apj, 744,
  71, \dodoi{10.1088/0004-637X/744/1/71}

\bibitem[{{Kilpatrick} {et~al.}(2014){Kilpatrick}, {Bieging}, \&
  {Rieke}}]{2014ApJ...796..144K}
{Kilpatrick}, C.~D., {Bieging}, J.~H., \& {Rieke}, G.~H. 2014, \apj, 796, 144,
  \dodoi{10.1088/0004-637X/796/2/144}

\bibitem[{{Koralesky} {et~al.}(1998){Koralesky}, {Rudnick}, {Gotthelf}, \&
  {Keohane}}]{1998ApJ...505L..27K}
{Koralesky}, B., {Rudnick}, L., {Gotthelf}, E.~V., \& {Keohane}, J.~W. 1998,
  \apjl, 505, L27, \dodoi{10.1086/311604}

\bibitem[{{Krause} {et~al.}(2008){Krause}, {Birkmann}, {Usuda}, {Hattori},
  {Goto}, {Rieke}, \& {Misselt}}]{2008Sci...320.1195K}
{Krause}, O., {Birkmann}, S.~M., {Usuda}, T., {et~al.} 2008, Science, 320,
  1195, \dodoi{10.1126/science.1155788}

\bibitem[{{Maeda} {et~al.}(2009){Maeda}, {Uchiyama}, {Bamba}, {Kosugi},
  {Tsunemi}, {Helder}, {Vink}, {Kodaka}, {Terada}, {Fukazawa}, {Hiraga},
  {Hughes}, {Kokubun}, {Kouzu}, {Matsumoto}, {Miyata}, {Nakamura}, {Okada},
  {Someya}, {Tamagawa}, {Tamura}, {Totsuka}, {Tsuboi}, {Ezoe}, {Holt},
  {Ishida}, {Kamae}, {Petre}, \& {Takahashi}}]{2009PASJ...61.1217M}
{Maeda}, Y., {Uchiyama}, Y., {Bamba}, A., {et~al.} 2009, \pasj, 61, 1217,
  \dodoi{10.1093/pasj/61.6.1217}

\bibitem[{{McMullin} {et~al.}(2007){McMullin}, {Waters}, {Schiebel}, {Young},
  \& {Golap}}]{2007ASPC..376..127M}
{McMullin}, J.~P., {Waters}, B., {Schiebel}, D., {Young}, W., \& {Golap}, K.
  2007, in Astronomical Society of the Pacific Conference Series, Vol. 376,
  Astronomical Data Analysis Software and Systems XVI, ed. R.~A. {Shaw},
  F.~{Hill}, \& D.~J. {Bell}, 127

\bibitem[{{Morse} {et~al.}(2004){Morse}, {Fesen}, {Chevalier}, {Borkowski},
  {Gerardy}, {Lawrence}, \& {van den Bergh}}]{2004ApJ...614..727M}
{Morse}, J.~A., {Fesen}, R.~A., {Chevalier}, R.~A., {et~al.} 2004, \apj, 614,
  727, \dodoi{10.1086/423709}

\bibitem[{{Orlando} {et~al.}(2022){Orlando}, {Wongwathanarat}, {Janka},
  {Miceli}, {Nagataki}, {Ono}, {Bocchino}, {Vink}, {Milisavljevic}, {Patnaude},
  \& {Peres}}]{2022A&A...666A...2O}
{Orlando}, S., {Wongwathanarat}, A., {Janka}, H.~T., {et~al.} 2022, \aap, 666,
  A2, \dodoi{10.1051/0004-6361/202243258}

\bibitem[{{Patnaude} \& {Fesen}(2007)}]{2007AJ....133..147P}
{Patnaude}, D.~J., \& {Fesen}, R.~A. 2007, \aj, 133, 147,
  \dodoi{10.1086/509571}

\bibitem[{{Patnaude} \& {Fesen}(2009)}]{2009ApJ...697..535P}
---. 2009, \apj, 697, 535, \dodoi{10.1088/0004-637X/697/1/535}

\bibitem[{{Patnaude} \& {Fesen}(2014)}]{2014ApJ...789..138P}
---. 2014, \apj, 789, 138, \dodoi{10.1088/0004-637X/789/2/138}

\bibitem[{{Perley} \& {Butler}(2013)}]{2013ApJS..204...19P}
{Perley}, R.~A., \& {Butler}, B.~J. 2013, \apjs, 204, 19,
  \dodoi{10.1088/0067-0049/204/2/19}

\bibitem[{{Reed} {et~al.}(1995){Reed}, {Hester}, {Fabian}, \&
  {Winkler}}]{1995ApJ...440..706R}
{Reed}, J.~E., {Hester}, J.~J., {Fabian}, A.~C., \& {Winkler}, P.~F. 1995,
  \apj, 440, 706, \dodoi{10.1086/175308}

\bibitem[{{Reichart} \& {Stephens}(2000)}]{2000ApJ...537..904R}
{Reichart}, D.~E., \& {Stephens}, A.~W. 2000, \apj, 537, 904,
  \dodoi{10.1086/309073}

\bibitem[{{Rosenberg}(1970)}]{1970MNRAS.151..109R}
{Rosenberg}, I. 1970, \mnras, 151, 109, \dodoi{10.1093/mnras/151.1.109}

\bibitem[{{Sano} \& {Fukui}(2021)}]{2021Ap&SS.366...58S}
{Sano}, H., \& {Fukui}, Y. 2021, \apss, 366, 58,
  \dodoi{10.1007/s10509-021-03960-4}

\bibitem[{{Sato} {et~al.}(2018){Sato}, {Katsuda}, {Morii}, {Bamba}, {Hughes},
  {Maeda}, {Ishida}, \& {Fraschetti}}]{2018ApJ...853...46S}
{Sato}, T., {Katsuda}, S., {Morii}, M., {et~al.} 2018, \apj, 853, 46,
  \dodoi{10.3847/1538-4357/aaa021}

\bibitem[{{Thorstensen} {et~al.}(2001){Thorstensen}, {Fesen}, \& {van den
  Bergh}}]{2001AJ....122..297T}
{Thorstensen}, J.~R., {Fesen}, R.~A., \& {van den Bergh}, S. 2001, \aj, 122,
  297, \dodoi{10.1086/321138}

\bibitem[{{Trotter} {et~al.}(2017){Trotter}, {Reichart}, {Egger},
  {St{\'y}blov{\'a}}, {Paggen}, {Martin}, {Dutton}, {Reichart}, {Kumar},
  {Maples}, {Barlow}, {Berger}, {Foster}, {Frank}, {Ghigo}, {Haislip},
  {Heatherly}, {Kouprianov}, {LaCluyz{\'e}}, {Moffett}, {Moore}, {Stanley}, \&
  {White}}]{2017MNRAS.469.1299T}
{Trotter}, A.~S., {Reichart}, D.~E., {Egger}, R.~E., {et~al.} 2017, \mnras,
  469, 1299, \dodoi{10.1093/mnras/stx810}

\bibitem[{{Tuffs}(1986)}]{1986MNRAS.219...13T}
{Tuffs}, R.~J. 1986, \mnras, 219, 13, \dodoi{10.1093/mnras/219.1.13}

\bibitem[{{Uchiyama} \& {Aharonian}(2008)}]{2008ApJ...677L.105U}
{Uchiyama}, Y., \& {Aharonian}, F.~A. 2008, \apjl, 677, L105,
  \dodoi{10.1086/588190}

\bibitem[{{Vink} \& {Laming}(2003)}]{2003ApJ...584..758V}
{Vink}, J., \& {Laming}, J.~M. 2003, \apj, 584, 758, \dodoi{10.1086/345832}

\bibitem[{{Vink} {et~al.}(2022){Vink}, {Patnaude}, \&
  {Castro}}]{2022ApJ...929...57V}
{Vink}, J., {Patnaude}, D.~J., \& {Castro}, D. 2022, \apj, 929, 57,
  \dodoi{10.3847/1538-4357/ac590f}

\bibitem[{{Weil} {et~al.}(2020){Weil}, {Fesen}, {Patnaude}, {Raymond},
  {Chevalier}, {Milisavljevic}, \& {Gerardy}}]{2020ApJ...891..116W}
{Weil}, K.~E., {Fesen}, R.~A., {Patnaude}, D.~J., {et~al.} 2020, \apj, 891,
  116, \dodoi{10.3847/1538-4357/ab76bf}

\bibitem[{{Willingale} {et~al.}(2002){Willingale}, {Bleeker}, {van der Heyden},
  {Kaastra}, \& {Vink}}]{2002A&A...381.1039W}
{Willingale}, R., {Bleeker}, J.~A.~M., {van der Heyden}, K.~J., {Kaastra},
  J.~S., \& {Vink}, J. 2002, \aap, 381, 1039,
  \dodoi{10.1051/0004-6361:20011614}

\bibitem[{{Wright} {et~al.}(1999){Wright}, {Dickel}, {Koralesky}, \&
  {Rudnick}}]{1999ApJ...518..284W}
{Wright}, M., {Dickel}, J., {Koralesky}, B., \& {Rudnick}, L. 1999, \apj, 518,
  284, \dodoi{10.1086/307270}

\bibitem[{{Zirakashvili} {et~al.}(2014){Zirakashvili}, {Aharonian}, {Yang},
  {O{\~n}a-Wilhelmi}, \& {Tuffs}}]{2014ApJ...785..130Z}
{Zirakashvili}, V.~N., {Aharonian}, F.~A., {Yang}, R., {O{\~n}a-Wilhelmi}, E.,
  \& {Tuffs}, R.~J. 2014, \apj, 785, 130, \dodoi{10.1088/0004-637X/785/2/130}

\end{thebibliography}
\bibliographystyle{aasjournal}

\end{document}